\documentclass[journal]{IEEEtran}
\usepackage{array}
\usepackage{algorithm,algorithmic}
\usepackage{cite}
\usepackage{amsfonts}
\usepackage{amssymb}
\usepackage{amsmath}
\usepackage{soul}
\usepackage{xcolor}
\usepackage{mathrsfs}
\usepackage{stfloats}
\usepackage{bm}
\usepackage{graphicx}
\usepackage{subfigure}
\usepackage{color,soul}
\usepackage{cuted}
\usepackage{textcomp}
\usepackage{balance}
\usepackage{upgreek}
\usepackage{enumerate}
\usepackage{stfloats}

\IEEEaftertitletext{\vspace{-1.5\baselineskip}}

\newtheorem{theorem}{Theorem}

\newtheorem{proposition}{Proposition}
\newtheorem{lemma}{Lemma}
\newtheorem{remark}{Remark}

\begin{document}
\title{Explicit Performance Bound of Finite Blocklength Coded MIMO: Time-Domain versus Spatiotemporal Channel Coding}

\author{~Feng~Ye,~\IEEEmembership{Student~Member,~IEEE,}~Xiaohu~You,~\IEEEmembership{Fellow,~IEEE,}
~Jiamin~Li,~\IEEEmembership{Member,~IEEE,}~Chen~Ji~and~Chuan~Zhang,~\IEEEmembership{Member,~IEEE}
\thanks{This work was supported in part by the National Key R\&D Program of China under Grant 2021YFB2900300, by the Fundamental Research Funds for the Central Universities under Grant 2242022k60006, by the Major Key Project of PCL (PCL2021A01-2), and by the Postgraduate Research\&Practice Innovation Program of Jiangsu Province under Grant KYCX24\_0408. \emph{(Corresponding author: Xiaohu You.)}}
\thanks{Feng Ye, Xiaohu You, Jiamin Li and Chuan Zhang are with the National Mobile Communications Research Laboratory, Southeast University, Nanjing {\rm 210096}, China, and also with Purple Mountain Laboratories, Nanjing {\rm 211111}, China (e-mail: yefeng@seu.edu.cn; xhyu@seu.edu.cn; jiaminli@seu.edu.cn; chzhang@seu.edu.cn).}
\thanks{Chen Ji is with School of Information Science and Technology, Nantong University, Nantong {\rm 226019}, China. (e-mail: gwidjin@ntu.edu.cn)}
}

\maketitle

\begin{abstract}
	In the sixth generation (6G), ultra-reliable low-latency communications (URLLC) will be further developed to achieve $\text{TK}\upmu$ extreme connectivity.
	On the premise of ensuring the same rate and reliability, the spatial domain advantage of multiple-input multiple-output (MIMO) has the potential to further shorten the time-domain code length and is expected to be a key enabler for the realization of $\text{TK}\upmu$.
	Different coded MIMO schemes exhibit disparities in exploiting the spatial domain characteristics, so we consider two extreme MIMO coding schemes, namely, time-domain coding in which the codewords on multiple spatial channels are independent of each other, and spatiotemporal coding in which multiple spatial channels are jointly coded.
	By analyzing the statistical characteristics of information density and utilizing the normal approximation, we provide explicit performance bounds for finite blocklength coded MIMO under time-domain coding and spatiotemporal coding.
	It is found that, different from the phenomenon in time-domain coding where the performance declines as the blocklengths decrease, spatiotemporal coding can effectively compensate for the performance loss caused by short blocklengths by improving the spatial degrees of freedom (DoF).
	These results indicate that spatiotemporal coding can optimally exploit the spatial dimension advantages of MIMO systems, enabling extremely low error-rate communication under stringent blocklengths constraint.
\end{abstract}


\section{Introduction}
\IEEEPARstart{A}{s} one of the main communication scenarios of the fifth generation (5G) mobile communication, ultra-reliable low-latency communications (URLLC) is the basis for achieving mission-critical applications with strict requirements for end-to-end latency and reliability \cite{schulz2017latency}. 
Driven by the increasingly stringent requirements of new applications such as extended reality (XR), industrial automation and telemedicine \cite{she2021tutorial,tariq2020speculative}, the capabilities of URLLC in the sixth generation (6G) are expected to grow further. 
Some application scenarios (such as XR) need to provide high reliability, low latency and high data rate services for devices at the same time \cite{saad2019vision}. 
Moreover, latency will be reduced from the $\text{ms}$-order of 5G to the $\upmu \text{s}$-order of 6G, and reliability will be increased from 99.999\% to 99.9999\%.
To this end, we abbreviate the three key performance indicators (KPIs) that 6G needs to achieve as $\text{TK}\upmu$ extreme connectivity, i.e., 1 Tbps data rate, 1 Kbps/Hz spectral efficiency, and $\upmu$s-level latency for vast applications \cite{you2023toward}. 
The theoretical evaluation of the realization of these KPIs is of significant importance in facilitating the design of the appropriate architecture and key technologies for 6G URLLC transmission.

Since URLLC relies on shorter packet length and smaller transmission interval \cite{bennis2018ultrareliable}, the length of codewords will be substantially reduced to meet the stringent delay constraint.
Finite blocklength coding addresses the practical latency constraints by optimizing the trade-off between blocklength (latency), decoding error probability (reliability), and coding rate \cite{hesham2023finite}.
In \cite{polyanskiy2010channel} and \cite{polyanskiy2009dispersion}, the nonasymptotic upper and lower bounds and normal approximations on coding rate as a function of decoding error probability and blocklength at finite blocklength for the binary erasure channel (BEC), the binary symmetric channel (BSC), and the additive white Gaussian noise (AWGN) channel are summarized and analyzed. 
In a succession of subsequent research efforts, these results have been extended to other types of scalar channels, such as \cite{polyanskiy2011scalar,polyanskiy2013dispersion,ostman2018short,lancho2019single}. 
Obviously, in scalar channels, the amount of data information decreases with decreasing blocklength for given decoding error probability, and it evidently fails to meet the progressively increasing demands of 6G URLLC.

Multiple-input multiple-output (MIMO) technology introduces additional spatial domain degree of freedom (DoF) by deploying multiple antennas at both ends of the transceivers \cite{zhu2023movable}.
MIMO is considered to be a key technology for guaranteeing communication performance under low latency requirements in URLLC scenarios \cite{popovski2019wireless,ren2020joint,ostman2021urllc}. 
Despite the fact that the normal approximation is not as accurate as the saddlepoint method \cite{martinez2011saddlepoint,lancho2020saddlepoint,lancho2021finite} at low coding rates, its concise form makes it easier to intuitively analyze the relationship between various KPIs and theoretically illustrate the internal logic of MIMO supporting URLLC transmission.
There have been a series of research efforts to extend the normal approximation in \cite{polyanskiy2010channel} to MIMO systems. 
In \cite{yang2014quasi} and \cite{yang2014dispersion}, the coding rate of a quasi-static MIMO channel was described given blocklength and decoding error probability. 
\cite{collins2016dispersion} presented the expression of the channel dispersion in MIMO coherent fading channel, and \cite{qi2020high} presented a high-SNR normal approximation of the maximum coding rate in a non-coherent Rayleigh block-fading MIMO channel.
Given that the rate-latency-reliability relationship obtained from the aforementioned research is overly complicated to directly infer the effect of the number of antennas on the performance of a finite blocklength MIMO system, \cite{you2023closed} further derived the explicit solutions for the average maximal achievable rate in MIMO systems with respect to spatial DoF, decoding error probability and blocklength based on the conclusion of \cite{yang2014quasi}.
The results show that the spatial DoF (coding in spatial-temporal domain) can effectively compensate for the performance loss caused by the reduction of temporal DoF due to shortened codewords.
\cite{you20236g} provided the theoretical upper bound for spatiotemporal two-dimensional (2-D) channel coding and suggested that through spatiotemporal coding, transmission reliability and latency can be flexibly balanced in a variety of communication scenarios including URLLC.
We can find that the above finite blocklength theoretical studies on MIMO channels all adopt the joint spatiotemporal coding scheme with multiple antennas at the transmitter. 
However, the time-domain coding scheme with independent coding at each transmit antenna can also be used in practical MIMO coding. 
A spatiotemporal 2-D polar coding scheme was proposed in \cite{you2022spatiotemporal}, and simulation analysis reveals that spatiotemporal coding is capable of fully exploiting the spatial domain resources to achieve high data-rate performance under low latency constraint, while time-domain coding will not benefit from spatial resources in comparison.

To the best of our knowledge, the intuitive performance superiority of spatiotemporal coding over time-domain coding has not yet been thoroughly investigated from a theoretical point of view.
In order to comprehensively understand the benefits of spatiotemporal coding for URLLC transmission in MIMO systems, it is essential to derive explicit solutions for the achievable rates of MIMO systems employing spatiotemporal and time-domain coding schemes.
The theoretical tools in \cite{polyanskiy2010channel,polyanskiy2009dispersion,yang2014quasi,yang2014dispersion,collins2016dispersion,qi2020high,you2023closed,you20236g} enable one to characterize the maximal achievable rates of spatiotemporal coding and time-domain coding by analytical means, and it is of great significance to additionally find an explicit expression to identify how spatial DoF improves overall performance. 
The novel theoretical studies in finite blocklength coding have the potential to provide a new perspective into practical spatiotemporal coding design.

In this paper, through the analysis of achievability bounds and converse bounds, the normal approximation expressions for spatiotemporal coding and time-domain coding are presented. 
Subsequently, the explicit closed-form upper bounds on the maximal achievable rate of the system under different coding methods are obtained by resolving the statistical properties of the first and second moments of the information density.
Based on the derived explicit solution, we conduct an analysis of the performance advantages of spatiotemporal coding over time-domain coding from the perspectives of normalized maximal achievable rate and decoding error probability.
The main contributions of this paper are summarized as follows:

\begin{enumerate}
	\setlength{\itemsep}{0pt}
	\setlength{\parsep}{0pt}
	\setlength{\parskip}{0pt}
	\item The achievability bound and the converse bound are respectively employed to provide normal approximation results for spatiotemporal and time-domain coding. It is demonstrated that the normal approximation remains accurate even under the conditions of extremely short blocklength and low decoding error probability in our high signal-to-noise ratio (SNR) and massive MIMO scenarios.
	\item We derive compact and explicit approximations for the average maximal achievable rate of coded massive MIMO with high SNR under different encoding modes. When considering a quasi-static flat Rayleigh fading MIMO channel and the channel state information (CSI) is unknown at the transmitter and known at the receiver, the maximal coding rate of each link in time-domain coding is given by
	\begin{equation}
	\frac{{{{\bar{R}}}_{\text{TD}}}}{m}\approx \log \left( 1+\rho  \right)-\frac{1}{\sqrt{n}}\frac{{{\Phi }^{-1}}\left( \varepsilon  \right)}{\ln 2}.
	\end{equation}
	And for spatiotemporal coding, it is given by
	\begin{equation}
	\frac{{{{\bar{R}}}_{\text{ST}}}}{m}\approx \log \left( 1+\rho  \right)-\sqrt{\frac{1}{mn}}\frac{{{\Phi }^{-1}}\left( \varepsilon  \right)}{\ln 2}.
	\end{equation}
	We assume that the number of transmit antennas in the MIMO system is $L$, the number of receive antennas is $N$, and the spatial DoF is defined as $m\triangleq \min \left\{ L,N \right\}$. The blocklength is $n$, and the decoding error probability is $\varepsilon$. Here, $\rho$ represents the SNR, and ${{\Phi }^{-1}} \left( \cdot  \right)$ stands for the inverse of the Gaussian Q-function. This demonstrates that compared with time-domain coding, the Shannon capacity can be approximated by increasing the spatial DoF in spatiotemporal coding, thereby mitigating the performance degradation induced by latency reduction.
	\item After further transformation, with the coding rate of each link being ${{\bar{R}}}/{m}\;$, spatiotemporal coding can improve the reliability (decoding error probability) of the system by increasing the spatial DoF compared with time-domain coding
	\begin{equation}
	{{\varepsilon }^{\text{ST}}}\approx \Phi \left[ \left( \log \left( 1+\rho  \right)-\frac{{\bar{R}}}{m} \right)\sqrt{mn}\ln 2 \right].
	\end{equation}
	As the spatial DoF approaches infinity, a finite blocklength can attain an arbitrarily low decoding error probability via spatiotemporal coding.
\end{enumerate}

The remainder of the paper is organized as follows. In Section II, we give some preliminaries. The explicit performance bounds of finite blocklength coded MIMO in different coding modes are derived in Section III. Section IV analyzes the normalized maximal achievable rate and decoding error probability. Section V provides numerical results to reveal the advantages of spatiotemporal coding versus time-domain coding in finite blocklength coded MIMO.

{\emph{Notations}}: Upper case letters such as $X$ represent scalar random variables and their realizations are written in lower case letters, e.g., $x$.
The boldface upper case letters represent random vectors, e.g., $\textbf{X}$, and boldface lower case letters represent their realizations, e.g., $\textbf{x}$.
Upper case letters of special fonts are used to denote deterministic matrices (e.g., $\mathsf{X}$), random matrices (e.g., $\mathbb{X}$) and sets (e.g., $\mathcal{X}$).
A N-dimensional identity matrix is denoted by $\mathsf{I}_N$, and $\mathbb{C}^{M\times N}$ denotes complex matrices with dimension $M\times N$.
The notation $\left( \cdot \right) ^\text{H}$ denotes the conjugate transpose of a vector or matrix.
Moreover, we use $\text{tr}\left( \cdot \right)$ and $\det \left( \cdot \right)$ to denote the trace and determinant of a matrix, respectively.
The mean and variance of a random variable are illustrated by the operators $\mathbb{E}\left( \cdot \right)$ and $\text{Var}\left( \cdot \right)$.
Finally, $\mathcal{CN}(\mu,\sigma^2)$ denotes the circularly symmetric complex Gaussian distribution with mean $\mu$ and variance $\sigma^2$.


\section{Preliminaries}
In this section, we introduce the definitions of information density and binary hypothesis testing under the channel model fixing the output distribution (such as AWGN channel). 
Based on this, we provide the achievability bound and converse bound to facilitate the subsequent solution of the normal approximations for different coding schemes of the MIMO channel.
\subsection{Information Density}
Consider an abstract channel model defined by a triple: input and output measurable spaces $\mathcal{A}$ and $\mathcal{B}$ and a conditional probability measure ${{P}_{Y|X}}:\mathcal{A}\to \mathcal{B}$.
Denote a codebook with $M$ codewords by $\left\{ {{c}_{1}},\ldots ,{{c}_{M}} \right\}\subset \mathcal{A}$. 
A (possibly randomized) decoder is a random transformation ${{P}_{Z|Y}}:\mathcal{B}\to \left\{ 0,1,\ldots ,M \right\}$, where `0' denotes that the decoder chooses ``error".
The maximal error probability is
\begin{equation}
\varepsilon =\underset{m\in \left\{ 1,\ldots ,M \right\}}{\mathop{\max }}\,\left[ 1-{{P}_{Z|X}}\left( m|{{c}_{m}} \right) \right]. \nonumber
\end{equation}
A codebook with $M$ codewords and a decoder satisfies ${{P}_{Z|X}}\left( m|{{c}_{m}} \right)\ge 1-\varepsilon $ are called an $(M,\varepsilon )$-code.
For a joint distribution ${{P}_{{{X}}{{Y}}}}$ on ${{\mathcal{A}}}\times {{\mathcal{B}}}$, the information density can be expressed as
\begin{equation}
	\iota \left( {{x}};{{y}} \right)=\log \frac{d{{P}_{{{Y}}|{{X}={x}}}}}{d{{P}_{{{Y}}}}}\left( {{y}} \right).
\end{equation}

In consideration of the transmission process of encoded codewords with a length of $n$ (i.e., the blocklength is $n$), we take ${{\mathcal{A}}^{n}}$ and ${{\mathcal{B}}^{n}}$ to be the $n$-fold Cartesian products of alphabets $\mathcal{A}$ and $\mathcal{B}$, respectively. 
The channel is represented by a conditional probability sequence $\left\{{{P}_{{\mathbf{Y}}|{\mathbf{X}}}}:{\mathcal{A}^{n}}\to {\mathcal{B}^{n}} \right\}$.
An $(M,\varepsilon )$-code for $\left\{ {{\mathcal{A}}^{n}},{{\mathcal{B}}^{n}},{{P}_{{\mathbf{Y}}|{\mathbf{X}}}} \right\}$ is called an $(n,M,\varepsilon )$-code. 
Given the decoding error probability $\varepsilon$ and blocklength $n$, the maximal number of codewords that can be attained is expressed as
\begin{equation}
{{M}^{*}}\left( n,\varepsilon  \right)=\max \left\{ M:\exists \left( n,M,\varepsilon  \right)\text{-code} \right\}.
\end{equation}

\subsection{Binary Hypothesis Testing}
Consider a random variable $w$ defined on set $\mathcal{W}$ which is capable of taking probability measures $P$ or $Q$. 
A randomized test between these two distributions is defined by a random transformation ${{P}_{Z|W}}:\mathcal{W}\to \left\{ 0,1\right\}$, where 0 implies that the test chooses $Q$. 
The optimal achievable performance among those randomized tests is specified by the Neyman-Pearson function \cite{polyanskiy2010channel}
\begin{equation}
{{\beta }_{\alpha }}\left( P,Q \right)=\underset{{{P}_{Z|W}}:\sum\limits_{w\in \mathcal{W}}{P\left( w \right){{P}_{Z|W}}\left( 1|w \right)}\ge \alpha }{\mathop{\min }}\,Q\left( w \right){{P}_{Z|W}}\left( 1|w \right).
\end{equation}

The cost constraint for each codeword can be defined by specifying a subset $\mathcal{F} \subset \mathcal{A}$ of the permissible inputs. 
For an arbitrary $\mathcal{F} \subset \mathcal{A}$, we define an associated performance measure for the composite hypothesis test between ${{Q}_{{{Y}}}}$ and $\left\{ {{P}_{{{Y}}|{{X}={x}}}} \right\}_{{x} \in \mathcal{F}}$:
\begin{equation}
	{{\kappa }_{\tau }}\left( \mathcal{F},{{Q}_{Y}} \right)=\underset{{{P}_{Z|Y}}:{{\inf }_{x\in F}}{{P}_{Z|X}}\left( 1|x \right)\ge \tau }{\mathop{\inf }}\,{{Q}_{Y}}\left( y \right){{P}_{Z|Y}}\left( 1|y \right)
\end{equation}
When considering a communication process with codeword length $n$, the above two functions are denoted as ${\beta }_{\alpha }^n$ and ${\kappa }_{\tau }^n$.

\begin{lemma} 
	[\!\!\cite{polyanskiy2010channel}]
	For any $\gamma>0$,
	\begin{equation}\label{ht1}
		\alpha \le P\left[\frac{dP}{dQ}\ge \gamma \right] + \gamma {{\beta }_{\alpha }}\left( P,Q \right).
	\end{equation}
	At the same time,
	\begin{equation}\label{ht2}
		{{\beta }_{\alpha }}\left( P,Q \right) \le \frac{1}{\gamma_0}
	\end{equation}
	where $\gamma_0$ satisfies
	\begin{equation}\label{ht3}
		P\left[\frac{dP}{dQ}\ge \gamma_0 \right] \ge \alpha.
	\end{equation}
\end{lemma}\label{lemma1}

\begin{lemma} 
	[\!\!\cite{mann1996berry}, Berry-Esseen central-limit theorem (CLT)]
	Let $S_j, j=1,\dots,n$ be independent random variables with mean $\mu = \mathbb{E}\left[S_j\right]$, variance $\sigma^2_j=\text{Var}\left[S_j\right]$ and third absolute central moment ${{\theta }_{j}}=\mathbb{E}\left[ {{\left| {{S}_{j}}-{{\mu }_{j}} \right|}^{3}} \right]$. Then, for every $t\in \mathbb{R}$,
	\begin{equation}
		\left| P\left[ \frac{\sum\nolimits_{j=1}^{n}{\left( {{S}_{j}}-{{\mu }_{j}} \right)}}{\sqrt{\sum\nolimits_{j=1}^{n}{\sigma _{j}^{2}}}}\ge t \right]-Q\left( t \right) \right|\le \frac{6\sum\nolimits_{j=1}^{n}{{{\theta }_{j}}}}{{{\left( \sum\nolimits_{j=1}^{n}{\sigma _{j}^{2}} \right)}^{\frac{3}{2}}}}.
	\end{equation}
\end{lemma}\label{lemma2}

\subsection{Achievability and Converse Bounds}
\begin{lemma} 
	[\!\!\cite{polyanskiy2010channel}, Theorem 25]
	For any $0<\varepsilon<1$, there exists an $\left(M,\varepsilon\right)$-code with codewords chosen from $\mathcal{F} \subset \mathcal{A}$, satisfying
	\begin{equation}
		M\ge \underset{0<\tau <\varepsilon }{\mathop{\sup }}\,\underset{{{Q}_{Y}}}{\mathop{\sup }}\,\frac{{{\kappa }_{\tau }}\left( \mathcal{F},{{Q}_{Y}} \right)}{{{\sup }_{x\in \mathcal{F}}}{{\beta }_{1-\varepsilon +\tau }}\left( x,{{Q}_{Y}} \right)}.
	\end{equation}\label{lemma3}
\end{lemma}

\begin{lemma} 
	[\!\!\cite{polyanskiy2010channel}, Theorem 31]
	Every $\left(M,\varepsilon\right)$-code (maximal error probability) with codewords belonging to $\mathcal{F}$ satisfies
	\begin{equation}
		M\le \underset{{{Q}_{Y}}}{\mathop{\inf }}\,\underset{x\in \mathcal{F}}{\mathop{\sup }}\,\frac{1}{{{\beta }_{1-\varepsilon }}\left( x,{{Q}_{Y}} \right)}.
	\end{equation}\label{lemma4}
\end{lemma}

\section{Performance Bounds of MIMO Systems under Different Coding Modes}
In this section, we first present the MIMO channel model and analyze the difference between spatiotemporal coding and time-domain coding.
Subsequently, we proceed to solve the performance bounds that MIMO channels can achieve in spatiotemporal coding and time-domain coding, respectively.

\subsection{MIMO Channel Model}
We consider a memoryless quasi-static flat Rayleigh fading MIMO channel, wherein the random fading coefficients remain invariant throughout the duration of each codeword.
This is a typical assumption for URLLC where the blocklength is usually short enough.
The relationship between channel input and channel output can be expressed as

\begin{equation}
\mathbb{Y}=\mathbb{X}\mathbb{H}+\mathbb{W},
\end{equation}
where $\mathbb{X}\in {{\mathbb{C}}^{n\times L}}$ denotes the transmitted signal, $\mathbb{Y}\in {{\mathbb{C}}^{n\times N}}$ represents the corresponding received signal.
$\mathbb{W}\in {{\mathbb{C}}^{n\times N}}$ is the additive noise signal at the receiver, which comprises independent and identically distributed (i.i.d) $\mathcal{C}\mathcal{N}\left( 0,1 \right)$ entries.
The channel matrix $\mathbb{H}\in {{\mathbb{C}}^{L\times N}}$ contains random complex fading elements, with each element being an i.i.d $\mathcal{C}\mathcal{N}\left( 0,1 \right)$ Gaussian variable, and these elements remain constant over $n$ channel uses.
It is assumed that the transmitter has unknown CSI , while the receiver has perfect CSI, partly because there can be insufficient time for the receiver to feedback CSI during a URLLC transmission.
In a transmission process, $\mathbb{H}=\mathsf{H}$ is a deterministic channel, and we denote ${{\lambda }_{1}}\ge \ldots \ge {{\lambda }_{m}}$ as the eigenvalues of ${\mathsf{H}^{\text{H}}}\mathsf{H}$.
We take into account the capacity achieving input distribution described by the following lemma.

\begin{lemma} 
	[\!\!\cite{collins2016dispersion}, Proposition 3]
	The capacity achieving input distribution is uniquely $\mathbb{X}\in {{\mathbb{C}}^{n\times L}}$ where each entry $X_{i,j}$ has an i.i.d. $\mathcal{CN}\left(0,\rho/L\right)$ distribution.\label{lemma5}
\end{lemma}

\begin{remark}
It is reasonable to consider perfect CSI at the receiver in our system. 
On the one hand, we consider a quasi-static channel where the channel varies at intervals much larger than the codeword length. 
This is feasible in low mobility scenarios, such as automated factories, where there is typically sufficient time to transmit the pilots for attaining perfect CSI and subsequently transmit the data in the ensuing short blocks.
On the other hand, even in scenarios where the mobility is not low and the coherence time of the channel is close to the codeword length, some advanced channel estimation methods can still be utilized to predict CSI, given that the channel changes continuously in real environments. 
\end{remark}

\begin{figure}
	\centering
	\includegraphics[width=8cm]{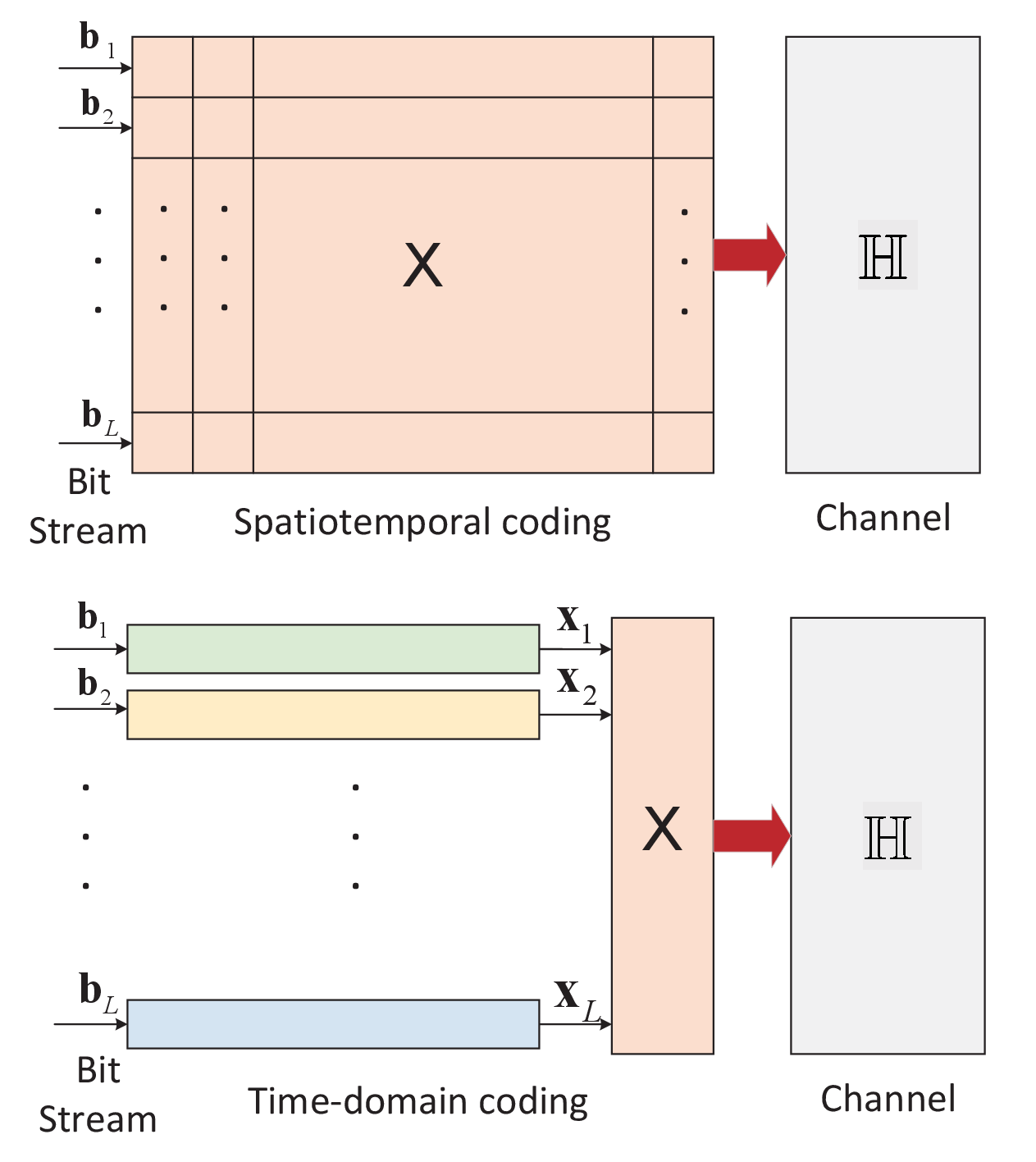}\\
	\caption{Different block diagrams of MIMO transmitting systems using spatiotemporal coding and time-domain coding.}\label{figure1}
\end{figure}

For channel coding in MIMO systems, two coding methods with different utilization of spatial DoF, spatiotemporal coding and time-domain coding, can be considered. 
The difference between the two methods is illustrated in Fig. \ref{figure1}.
In spatiotemporal coding, each bit stream is jointly encoded within both the time and spatial domain, thereby generating a complete codeword $\mathsf{X}$.
Then, the corresponding signals of the codeword in different spatial domains are sent out through $L$ transmit antennas.
In time domain coding, $L$ bit streams are encoded in the time domain. 
The codewords ${{\mathbf{x}}_{1}},\ldots ,{{\mathbf{x}}_{L}}$ corresponding to each column of the resulting transmitted signal $\mathsf{X}$ are independent of each other.
Thereafter, the time-domain codewords are dispatched through $L$ transmit antennas.

For the purpose of analyzing the maximal achievable performance of the MIMO system under different coding modes, the performance bound of the finite blocklength coded MIMO under a deterministic channel can be analyzed with respect to the channel $\mathsf{H}$.
Subsequently, the ergodic performance of the system under a random channel can be determined. 
It is worth noting that under the single-input single-output channel, the assumption of a quasi-static channel leads to the main error originating from the channel being in deep fading, which makes the maximal coding rate converge to the outage capacity. 
However, in massive MIMO channels, the channels experience different fading at different antennas, and the channels exhibit ergodicity even under the assumption of quasi-static channels.  
Therefore, it is appropriate to characterize the maximal achievable performance of a quasi-static MIMO channel in terms of ergodic capacity.
In the following, we will derive the explicit performance bound of finite blocklength coded MIMO from two different coding methods: spatiotemporal coding and time-domain coding.

\subsection{Spatiotemporal Coding}
For MIMO channels, we consider an optimal $\left( n,M_\text{ST}^{*},\varepsilon  \right)$-code, in which the codeword $\mathsf{X}$ is obtained through spatiotemporal coding and cannot be decomposed. 
Considering the imperfect CSI at the transmitter and capacity achieving input distribution as described in Lemma \ref{lemma5}, the input codeword is required to satisfy the power constraint $\mathcal{F}^\text{ST}=\left\{\mathsf{X}\in {\mathbf{C}^{n\times L}}:{{\left\| {\mathsf{X}_{\cdot ,i}} \right\|}^{2}}={n\rho }/{L}, i=1,\dots,L\right\}$.
The probability and conditional probability of the received signal are satisfied
\begin{align}
{{P}_{\mathbb{Y}\mathbb{H}}}=&{{P}_{\mathbb{H}}}\times {{P}_{\mathbb{Y}|\mathbb{H}}}, \\
{{P}_{\mathbb{Y}\mathbb{H}|\mathbb{X}}}=&{{P}_{\mathbb{H}}}\times {{P}_{\mathbb{Y}|\mathbb{X}\mathbb{H}}}.
\end{align}

According to \cite{yang2014quasi} and \cite{zhang2024mutual}, and by leveraging the probability density function of the Gaussian matrix, the information density of the MIMO channel $\mathsf{H}$ can be formulated as
\begin{align}
	&\iota\left( \mathsf{X};\mathbb{Y}|\mathsf{H} \right)  =\log \frac{d{{P}_{\mathbb{Y}\mathbb{H}|\mathbb{X}=\mathsf{X}}}}{d{{P}_{\mathbb{Y}\mathbb{H}}}} \nonumber \\
	&=\log \frac{d{{P}_{\mathbb{Y}|\mathbb{X}=\mathsf{X},\mathbb{H}=\mathsf{H}}}}{d{{P}_{\mathbb{Y}|\mathbb{H}=\mathsf{H}}}} \nonumber \\ 
	& =n\log \det \left( {\mathsf{I}_{N}}+\frac{1}{n}{\mathsf{H}^{\text{H}}}\mathsf{Q}_{X}\mathsf{H} \right) -\text{tr}\left[{{\mathbb{W}}^{\text{H}}}\mathbb{W}\right]\log e \nonumber \\
	& \quad +\text{tr}\left[ {{{\left( \mathsf{X}\mathsf{H}+\mathbb{W} \right)}^{\text{H}}}\left( \mathsf{X}\mathsf{H}+\mathbb{W} \right)}\left({{\mathsf{I}_{N}}+\frac{1}{n}{\mathsf{H}^{\text{H}}}\mathsf{Q}_{X}\mathsf{H}}\right)^{-1}\right]\log e \nonumber \\ 
	& =n\log \det \left( {\mathsf{I}_{N}}+\frac{1}{n}{\mathsf{H}^{\text{H}}}\mathsf{Q}_{X}\mathsf{H} \right) + \text{tr}\bigg[\left({{\mathsf{I}_{N}}+\frac{1}{n}{\mathsf{H}^{\text{H}}}\mathsf{Q}_{X}\mathsf{H}}\right)^{-1} \nonumber \\  
	& \quad \left({-\frac{1}{n}{\mathsf{H}^{\text{H}}}\mathsf{Q}_{X}\mathsf{H}{{\mathbb{W}}^{\text{H}}}\mathbb{W}+{{\mathbb{W}}^{\text{H}}}\mathsf{X}\mathsf{H}+{\mathsf{H}^{\text{H}}}{\mathsf{X}^{\text{H}}}\mathbb{W}+\mathsf{H}^{\text{H}}{\mathsf{X}^{\text{H}}}\mathsf{X}\mathsf{H}}\right)\bigg]\log{e},
\end{align}
where $\mathsf{Q}_{X}=\mathbb{E}\left[\mathbb{X}^\text{H}\mathbb{X}\right]=n\frac{\rho}{L}\mathsf{I}_L$.

Due to the spherical symmetry of both $\mathcal{F}^\text{ST}$ (equivalent to ${{P}_{{\mathbb{Y}}|{\mathbb{X}=\mathsf{X},\mathbb{H}=\mathsf{H}}}}$) and capacity achieving output distribution
\begin{equation}
	{{P}_{{\mathbb{Y}}|\mathbb{H}=\mathsf{H}}}= \prod\limits_{j=1}^{n} {\mathcal{CN}\left(\mathbf{0},{{\mathsf{I}_{N}}+\frac{1}{n}{\mathsf{H}^{\text{H}}}\mathsf{Q}_{X}\mathsf{H}}\right)}, \nonumber
\end{equation}
we can obtain that ${{\beta }_{\alpha }^n}\left( {{P}_{{\mathbb{Y}}|{\mathbb{X}=\mathsf{X},\mathbb{H}=\mathsf{H}}}},{{P}_{{\mathbb{Y}}|\mathbb{H}=\mathsf{H}}} \right)$ has spherical symmetry and does not depend on $\mathsf{X} \in \mathcal{F}^\text{ST}$. 
Therefore, given a codeword $\mathsf{X}^\text{H}\mathsf{X}=n\frac{\rho}{L}\mathsf{I}_L$ satisfying the power constraint $\mathcal{F}^\text{ST}$, the information density can be further transformed into
\begin{align}
	&\iota\left( \mathsf{X};\mathbb{Y}|\mathsf{H} \right)\nonumber\\
	&=n\log \det \left( {\mathsf{I}_{N}}+\frac{\rho}{L}{\mathsf{H}^{\text{H}}}\mathsf{H} \right)+ \text{tr}\bigg[\left({{\mathsf{I}_{N}}+\frac{\rho}{L}{\mathsf{H}^{\text{H}}}\mathsf{H}}\right)^{-1}\nonumber \\
	&\qquad \left({-\frac{\rho}{L}{\mathsf{H}^{\text{H}}}\mathsf{H}{{\mathbb{W}}^{\text{H}}}\mathbb{W}+{{\mathbb{W}}^{\text{H}}}\mathsf{X}\mathsf{H}+{\mathsf{H}^{\text{H}}}{\mathsf{X}^{\text{H}}}\mathbb{W}+n\frac{\rho}{L}\mathsf{H}^{\text{H}}\mathsf{H}}\right) \bigg]\log{e} \nonumber \\
	& = n\sum\limits_{i=1}^{m}{\log \left( 1+\frac{\rho }{L}{{\lambda }_{i}} \right)} \nonumber \\
	&\quad +\text{tr}\bigg[ n{\mathsf{I}_{N}}-\left({\frac{\rho }{L}{\mathsf{H}^{\text{H}}}\mathsf{H}{{\mathbb{W}}^{\text{H}}}\mathbb{W}-{{\mathbb{W}}^{\text{H}}}\mathsf{XH}-{\mathsf{H}^{\text{H}}}{\mathsf{X}^{\text{H}}}\mathbb{W}+n{\mathsf{I}_{N}}}\right) \nonumber \\
	& \qquad \qquad \qquad \left({{\mathsf{I}_{N}}+\frac{\rho }{L}{\mathsf{H}^{\text{H}}}\mathsf{H}}\right)^{-1} \bigg]\log e \nonumber \\
	& = \sum\limits_{j=1}^{n}{\sum\limits_{i=1}^{m}{\left[ \log \left( 1+\frac{\rho }{L}{{\lambda }_{i}} \right)+\left( 1-\frac{{{\left| \sqrt{\frac{\rho }{L}{{\lambda }_{i}}}{{W}_{i,j}}-1 \right|}^{2}}}{1+\frac{\rho }{L}{{\lambda }_{i}}} \right)\log e \right]}},
\end{align}
where ${{W}_{i,j}}\sim \mathcal{CN}\left( 0,1 \right)$.

\begin{proposition}
	Using the Berry-Esseen CLT, the Neyman-Pearson function for the MIMO channel under spatiotemporal coding can be expressed as
	\begin{align}\label{ht4}
		&{{\beta }_{\alpha}^n}\left( {{P}_{{\mathbb{Y}}|{\mathbb{X}=\mathsf{X},\mathbb{H}=\mathsf{H}}}},{{P}_{{\mathbb{Y}}|\mathbb{H}=\mathsf{H}}} \right)\nonumber \\
		&\qquad=-n{{C}^\text{ST}}\left( \mathsf{H} \right)+\sqrt{n{{V}^\text{ST}}\left( \mathsf{H} \right)}\Phi^{-1}\left(1-\alpha\right) + \mathcal{O}\left(\log n\right),
	\end{align}
	where the channel capacity and channel dispersion of the MIMO channel with spatiotemporal codewords in a channel use are expressed as
	\begin{equation}\label{c1}
		{{C}^\text{ST}}\left( \mathsf{H} \right)=\frac{1}{n}\mathbb{E}\left[ \iota\left( \mathsf{X};\mathbb{Y}|\mathsf{H} \right) \right]=\sum\limits_{i=1}^{m}{\log \left( 1+\frac{\rho }{L}{{\lambda }_{i}} \right)}=\sum\limits_{i=1}^{m}{{{C}_{i}}}
	\end{equation}
	and
	\begin{align}\label{v1}
		{{V}^\text{ST}}\left( \mathsf{H} \right)&=\frac{1}{n}\text{Var}\left[ \iota\left( \mathsf{X};\mathbb{Y}|\mathsf{H} \right) \right]\nonumber \\
		&=\sum\limits_{i=1}^{m}{\left[ 1-\frac{1}{{{\left( 1+\frac{\rho }{L}{{\lambda }_{i}} \right)}^{2}}} \right]}{{\log }^{2}}e=\sum\limits_{i=1}^{m}{{{V}_{i}}}.
	\end{align}
	We define that
	\begin{align}
		{{C}_{i}}&\triangleq \log \left( 1+\frac{\rho }{L}{{\lambda }_{i}} \right),  \\
		{{V}_{i}}&\triangleq \left[ 1-\frac{1}{{{\left( 1+\frac{\rho }{L}{{\lambda }_{i}} \right)}^{2}}} \right]{{\log }^{2}}e. 
	\end{align}
	
	{\emph{Proof: }} See Appendix \ref{ap1}.
\end{proposition}

\subsubsection{Achievability}
\begin{lemma} 
	[\!\!\cite{collins2018coherent}, Lemma 18]
	For all $\tau \in \left[ 0,1 \right]$ and a $\tau$-dependent constant $K_1>0$, we have
	\begin{equation}
		{\kappa }_{\tau }^n \left(\mathcal{F}^n,P_{\mathbf{Y}}\right) \ge K_1\left({\tau}\right).
	\end{equation}\label{lemma6}
\end{lemma}
To conclude the proof of achievability we apply Lemma \ref{lemma3} with $\tau = B/\sqrt{n}$ ($B$ is defined in Appendix A) and find that
\begin{align} \label{ache1}
&\log M_\text{ST}^{*}\left( n,\varepsilon  \right)\nonumber \\
&\ge \log {\kappa_\tau^n\left(\mathcal{F}^\text{ST},{{P}_{{\mathbb{Y}}|\mathbb{H}=\mathsf{H}}}\right)}-\log{\beta_{1-\varepsilon+\tau}^n\left( {{P}_{{\mathbb{Y}}|{\mathbb{X}=\mathsf{X},\mathbb{H}=\mathsf{H}}}},{{P}_{{\mathbb{Y}}|\mathbb{H}=\mathsf{H}}} \right)} \nonumber \\
&=n{{C}^\text{ST}}\left( \mathsf{H} \right)-\sqrt{n{{V}^\text{ST}}\left( \mathsf{H} \right)}\Phi^{-1}\left(\varepsilon-\tau \right)+\log \kappa_\tau^n+ \mathcal{O}\left(\log n\right) \nonumber \\
&\overset{\left( a \right)}{\mathop{\ge }} n{{C}^\text{ST}}\left( \mathsf{H} \right)-\sqrt{n{{V}^\text{ST}}\left( \mathsf{H} \right)}\Phi^{-1}\left(\varepsilon \right) + \mathcal{O}\left(\log n\right).
\end{align}
where $\left(a\right)$ makes use of Taylor's theorem and Lemma \ref{lemma6}.

\subsubsection{Converse}
According to Lemma \ref{lemma4}, we can obtain that the spatiotemporal  ($n,M_\text{ST}^{*},\varepsilon$)-code with codewords belonging to $\mathcal{F}^\text{ST}$ should satisfy
\begin{align}\label{con1}
\log M_\text{ST}^{*}\left( n,\varepsilon  \right) &\le -\log{\beta_{1-\varepsilon}^n\left( {{P}_{{\mathbb{Y}}|{\mathbb{X}=\mathsf{X},\mathbb{H}=\mathsf{H}}}},{{P}_{{\mathbb{Y}}|\mathbb{H}=\mathsf{H}}} \right)} \nonumber \\
&= n{{C}^\text{ST}}\left( \mathsf{H} \right)-\sqrt{n{{V}^\text{ST}}\left( \mathsf{H} \right)}\Phi^{-1}\left(\varepsilon \right) + \mathcal{O}\left(\log n\right).
\end{align}

\subsubsection{Normal Approximation}
The normal approximation for the maximal number of codewords follows simply by combining (\ref{ache1}) with (\ref{con1}):
\begin{equation}\label{normal1}
\log M_\text{ST}^{*}\left( n,\varepsilon  \right)\approx n{{C}^\text{ST}}\left( \mathsf{H} \right)-\sqrt{n{{V}^\text{ST}}\left( \mathsf{H} \right)}{{\Phi }^{-1}}\left( \varepsilon  \right).
\end{equation}
Then, for the channel matrix $\mathsf{H}$, the maximal coding rate should be
\begin{align}\label{r1}
	R_\text{ST}^{*}\left( n,\varepsilon  \right) &= \frac{\log M_\text{ST}^{*}\left( n,\varepsilon  \right)}{n} \nonumber \\
	& \approx {{C}^\text{ST}}\left( \mathsf{H} \right)-\sqrt{\frac{{{V}^\text{ST}\left( \mathsf{H} \right)}}{n}}{{\Phi }^{-1}}\left( \varepsilon  \right).
\end{align}

\begin{theorem}\label{the1}
For a quasi-static flat Rayleigh fading MIMO channel, the normal approximation holds with high accuracy provided that one of the following four conditions is satisfied:
\begin{itemize}
	\item Large Blocklength;
	\item High Decoding Error Probability;
	\item High SNR;
	\item Large Spatial DoF.
\end{itemize}

{\emph{Proof: }} See Appendix \ref{ap2}.
\end{theorem}

From Theorem \ref{the1}, it can be clearly observed that in the URLLC scenario of 6G, which demands extremely low latency and extremely high reliability, the accuracy of the normal approximation is compromised.
Nevertheless, with the augmentation of spatial DoF and SNR, the normal approximation is capable of attaining acceptable accuracy.
Consequently, the normal approximation formulation is applicable to precisely characterize the system performance within our hypothesized high SNR massive MIMO scenario.

\subsubsection{Maximal Achievable Rate}
In the case of the MIMO random channel matrix $\mathbb{H}$, the maximal achievable rate of the system ${{{\bar{R}}}_\text{ST}}$ can be solved by
\begin{align}
{{{\bar{R}}}_\text{ST}} &= \mathbb{E}\left[{{C}^\text{ST}}\left( \mathbb{H} \right)-{\sqrt{\frac{{{V}^\text{ST}}\left( \mathbb{H} \right)}{n}\;}}{{\Phi }^{-1}}\left( \varepsilon  \right) \right] \nonumber \\
& \overset{\left( a \right)}{\mathop{\approx}}\, \mathbb{E}\left[{{C}^\text{ST}}\left( \mathbb{H} \right)\right]-{\sqrt{\frac{\mathbb{E}\left[{{{V}^\text{ST}}\left( \mathbb{H} \right)}\right]}{n}\;}}{{\Phi }^{-1}}\left( \varepsilon  \right) 
\end{align}
where $\left(a\right)$ is due to the fact that in the case of high per-antenna SNR, i.e., ${\rho }/{L}\;\gg 1$, the variance of the channel dispersion tends to zero \cite{you2023closed}
\begin{align}
\text{Var}\left[ {{V}^\text{ST}}\left( \mathbb{H} \right) \right]\xrightarrow{{\rho }/{L}\;\gg 1} & \xi\left[\frac{L\left(N-L\right)}{8N}+\frac{L\left(L-N\right)}{8N}\right]^2\nonumber \\
&-\left[\frac{N\left(N-L\right)}{8L}+\frac{N\left(L-N\right)}{8L}\right]^2=0,
\end{align}
where $\xi<1$.
It is demonstrated that the variance of the channel dispersion is negligible, thereby allowing the channel dispersion to be approximately regarded as a deterministic value.

When ${\rho }/{L}\;\gg 1$, the approximated expectations of channel capacity and channel dispersion are given in \cite{you2023closed}
\begin{align}
\mathbb{E}\left[ {{C}^\text{ST}}\left( \mathbb{H} \right) \right]&\approx m\log \left( 1+\rho  \right), \label{c2} \\
\mathbb{E}\left[ {{V}^\text{ST}}\left( \mathbb{H} \right) \right]&\approx m{{\log }^{2}e}. \label{v2}
\end{align}
Therefore, the upper bound of the maximal achievable rate of the MIMO channel under spatiotemporal coding can be expressed as
\begin{equation}\label{r2}
	{{{\bar{R}}}_\text{ST}} \xrightarrow{{\rho }/{L}\;\gg 1}m\log \left( 1+\rho  \right)-\sqrt{\frac{m}{n}}\frac{{{\Phi }^{-1}}\left( \varepsilon  \right)}{\ln 2}  .
\end{equation}

\subsection{Time-Domain Coding}
In the case of time-domain coding, $L$ time-domain transmitted codewords ${{\mathbf{x}}_{1}},\ldots ,{{\mathbf{x}}_{L}}$ are utilized and their formations are mutually independent.
The transmitted signal can be expressed as $\mathsf{X}=\left[ {{\mathbf{x}}_{1}},\ldots ,{{\mathbf{x}}_{L}} \right]$. 
Given the requirement to apply the same codeword power constraints as in spatiotemporal coding, the codewords for time-domain coding need to be selected from the set ${{\mathcal{F}}^\text{TD}}=\left\{ {{\mathbf{x}}_{i}}\in {{\mathbb{C}}^{n}}:{{\left\| {{\mathbf{x}}_{i}} \right\|}^{2}}=n\rho/L,i=1,\dots,L \right\}$. 
In this case, the information density of MIMO channel $\mathsf{H}$ can also be expressed as (\ref{in_td}), which is shown in the botton of next page.
\begin{figure*}[!b]
	\normalsize
	\hrulefill
		\begin{align}\label{in_td}
			\iota \left( \mathsf{X};\mathbb{Y}|\mathsf{H} \right)
			& =n\log \det \left( {\mathsf{I}_{N}}+\frac{\rho}{L}{\mathsf{H}^{\text{H}}}\mathsf{H} \right)+  \text{tr}\left[\left({{\mathsf{H}^{\text{H}}}{\mathsf{X}^{\text{H}}}\mathsf{X}\mathsf{H} -\frac{\rho}{L}{\mathsf{H}^{\text{H}}}\mathsf{H}{{\mathbb{W}}^{\text{H}}}\mathbb{W}+{{\mathbb{W}}^{\text{H}}}\mathsf{X}\mathsf{H}+{\mathsf{H}^{\text{H}}}{\mathsf{X}^{\text{H}}}\mathbb{W}}\right) \left({{\mathsf{I}_{N}}+\frac{\rho}{L}{\mathsf{H}^{\text{H}}}\mathsf{H}}\right)^{-1} \right]\log{e} \nonumber \\
			& \overset{\left( a \right)}{\mathop{=}}\,n\log \det \left({\mathsf{I}_{N}}+\frac{\rho}{L}{\mathsf{H}^{\text{H}}}\mathsf{H} \right)- \text{tr}\left[{\frac{\rho}{L}\mathsf{V}\mathsf{\Lambda}^2{\mathsf{V}^{\text{H}}}{\mathbb{W}^{\text{H}}}\mathbb{W}}\left({{\mathsf{I}_{N}}+\frac{\rho}{L}\mathsf{V}\mathsf{\Lambda}^2{\mathsf{V}^{\text{H}}}}\right)^{-1} \right]\log e \nonumber \\ 
			& \quad + \text{tr}\left[\left({\mathsf{V}{{\Lambda }^{\text{H}}}{\mathsf{U}^{\text{H}}}{\mathsf{X}^{\text{H}}}\mathsf{X}\mathsf{U}\Lambda {\mathsf{V}^{\text{H}}}+{{\mathbb{W}}^{\text{H}}}\mathsf{X}\mathsf{U}\Lambda {\mathsf{V}^{\text{H}}}+\mathsf{V}{{\Lambda }^{\text{H}}}{\mathsf{U}^{\text{H}}}{\mathsf{X}^{\text{H}}}\mathbb{W}}\right)\left({{\mathsf{I}_{N}}+\frac{\rho}{L}\mathsf{V}\mathsf{\Lambda}^2{\mathsf{V}^{\text{H}}}}\right)^{-1} \right]\log e  \nonumber \\ 
			& \overset{\left( b \right)}{\mathop{=}}\,n\log \det \left({\mathsf{I}_{N}}+\frac{\rho}{L}{\mathsf{H}^{\text{H}}}\mathsf{H} \right)+\text{tr}\left[ \left({{\mathsf{\Lambda }^{\text{H}}}{{{\tilde{\mathsf{X}}}}^{\text{H}}}\tilde{\mathsf{X}}\mathsf{\Lambda}+{{{\tilde{\mathbb{W}}}}^{\text{H}}}\tilde{\mathsf{X}}\mathsf{\Lambda}+{\mathsf{\Lambda }^{\text{H}}}{{{\tilde{\mathsf{X}}}}^{\text{H}}}\tilde{\mathbb{W}}-\frac{\rho}{L}\mathsf{\Lambda}^2{{{\tilde{\mathbb{W}}}}^{\text{H}}}\tilde{\mathbb{W}}}\right)\left({{{\mathsf{I}_{N}}+\frac{\rho}{L}\mathsf{\Lambda}^2 }}\right)^{-1} \right]\log e \nonumber \\ 
			& =n\sum\limits_{i=1}^{m}{\log \left( 1+\frac{\rho}{L}{{\lambda }_{i}} \right)}+\sum\limits_{i=1}^{m}{\left\{ \text{tr}\left[ \frac{{{\lambda}_{i}}\mathbf{\tilde{x}}_{i}^{\text{H}}{{{\mathbf{\tilde{x}}}}_{i}}+\sqrt{{{\lambda }_{i}}}{{{\mathbf{\tilde{W}}}}_{i}}^{\text{H}}{{{\mathbf{\tilde{x}}}}_{i}}+\sqrt{{{\lambda }_{i}}}\mathbf{\tilde{x}}_{i}^{\text{H}}{{{\mathbf{\tilde{W}}}}_{i}}-\frac{\rho}{L}{{\lambda }_{i}}{{{\mathbf{\tilde{W}}}}_{i}}^{\text{H}}{{{\mathbf{\tilde{W}}}}_{i}}}{1+\frac{\rho}{L}{{\lambda }_{i}}}\right]\log e \right\} } \nonumber \\ 
			& =\sum\limits_{i=1}^{m}{\log \frac{d{{P}_{{{{\mathbf{\tilde{Y}}}}_{i}}|{{{\mathbf{\tilde{X}}}}_{i}}={{{\mathbf{\tilde{x}}}}_{i}},{{\Lambda }_{i}}=\sqrt{{{\lambda }_{i}}}}}}{d{{P}_{{{{\mathbf{\tilde{Y}}}}_{i}}|{{\Lambda }_{i}}=\sqrt{{{\lambda }_{i}}}}}}}=\sum\limits_{i=1}^{m}{{{\iota}_{i}}\left( {{{\mathbf{\tilde{x}}}}_{i}};{{{\mathbf{\tilde{Y}}}}_{i}}|{{\lambda }_{i}} \right)}  .
		\end{align}
\end{figure*}
(a) is to perform singular value decomposition (SVD) on matrix $\mathsf{H}=\mathsf{U}\mathsf{\Lambda} {\mathsf{V}^{\text{H}}}$, where $\mathsf{\Lambda}\in \mathbb{C}^{L\times N}$ is a diagonal matrix formed by diagonal elements $\sqrt{{{\lambda }_{1}}},\ldots ,\sqrt{{{\lambda }_{m}}}$, $\mathsf{U}\in \mathbb{C}^{L\times L}$ and $\mathsf{V}\in \mathbb{C}^{N\times N}$ are unitary matrices. 
In (b), we have $\tilde{\mathbb{Y}}=\tilde{\mathsf{X}}\mathsf{\Lambda} +\tilde{\mathbb{W}}$, where $\tilde{\mathbb{Y}}=\tilde{\mathbb{Y}}\mathsf{V}=\left[ {{{\mathbf{\tilde{Y}}}}_{1}},\ldots ,{{{\mathbf{\tilde{Y}}}}_{N}} \right]$, $\tilde{\mathsf{X}}=\mathsf{X}\mathsf{U}=\left[ {{{\mathbf{\tilde{x}}}}_{1}},\ldots ,{{{\mathbf{\tilde{x}}}}_{L}} \right]$ and $\tilde{\mathbb{W}}=\mathbb{W}\mathsf{V}=\left[ {{{\mathbf{\tilde{W}}}}_{1}},\ldots ,{{{\mathbf{\tilde{W}}}}_{N}} \right]$.
Since ${{\mathbf{x}}_{1}},\ldots ,{{\mathbf{x}}_{L}}$ are statistically independent,  it follows that when multiplied by the unitary matrix, ${{\mathbf{\tilde{x}}}_{1}},\ldots ,{{\mathbf{\tilde{x}}}_{L}}$ are also statistically independent and the codewords also obey the power constraint ${{\mathcal{F}}^\text{TD}}$ \cite{yang2014quasi}. 

The above formula reveals that the maximal number of codewords is acquired by analyzing $m$ independent link composed of codewords ${{\mathbf{\tilde{x}}}_{1}},\ldots ,{{\mathbf{\tilde{x}}}_{m}}$, which is equivalent to the maximal number of codewords obtained from the analysis of codewords ${{\mathbf{x}}_{1}},\ldots ,{{\mathbf{x}}_{L}}$.
If we take $m$ optimal $\left( n,M_{i}^{*},\varepsilon  \right)$-codes for $m$ links, we can get an $\left( n,\prod\nolimits_{i=1}^{m}{M_{i}^{*}},1-{{\left( 1-\varepsilon  \right)}^{m}} \right)$-code for a MIMO channel.
Consequently, in order to compute the maximal achievable performance of the MIMO system in time-domain coding, it is essential to calculate the maximal number of codewords that can be achieved by each codeword, which requires calculation based on the information density of the independent link formed by each codeword.

For link $i$, the probability and conditional probability of the received signal ${{{\mathbf{\tilde{Y}}}}_{i}}$ are ${{P}_{{{{\mathbf{\tilde{Y}}}}_{i}}|{{\Lambda }_{i}}=\sqrt{{{\lambda }_{i}}}}}$ and ${{P}_{{{{\mathbf{\tilde{Y}}}}_{i}}|{{{\mathbf{\tilde{X}}}}_{i}}={{{\mathbf{\tilde{x}}}}_{i}},{{\Lambda }_{i}}=\sqrt{{{\lambda }_{i}}}}}$.
Then, the information density of link $i$ is
\begin{align}
	&{{\iota}_{i}}\left( {{{\mathbf{\tilde{x}}}}_{i}};{{{\mathbf{\tilde{Y}}}}_{i}}|{{\lambda }_{i}} \right)=\log \frac{d{{P}_{{{{\mathbf{\tilde{Y}}}}_{i}}|{{{\mathbf{\tilde{X}}}}_{i}}={{{\mathbf{\tilde{x}}}}_{i}},{{\Lambda }_{i}}=\sqrt{{{\lambda }_{i}}}}}}{d{{P}_{{{{\mathbf{\tilde{Y}}}}_{i}}|{{\Lambda }_{i}}=\sqrt{{{\lambda }_{i}}}}}} \nonumber \\ 
	& =n\log \left( 1+\frac{\rho}{L}{{\lambda }_{i}}\right) \nonumber \\ 
	&\quad +\text{tr}\left[ \frac{{{\lambda }_{i}}\mathbf{\tilde{x}}_{i}^{\text{H}}{{{\mathbf{\tilde{x}}}}_{i}}+\sqrt{{{\lambda }_{i}}}{{{\mathbf{\tilde{W}}}}_{i}}^{\text{H}}{{{\mathbf{\tilde{x}}}}_{i}}+\sqrt{{{\lambda }_{i}}}\mathbf{\tilde{x}}_{i}^{\text{H}}{{{\mathbf{\tilde{W}}}}_{i}}-\frac{\rho}{L}{{\lambda }_{i}}{{{\mathbf{\tilde{W}}}}_{i}}^{\text{H}}{{{\mathbf{\tilde{W}}}}_{i}}}{1+\frac{\rho}{L}{{\lambda }_{i}}} \right]\log e.
\end{align}
To obtain the maximal number of codewords per link, the Neyman-Pearson function needs to be solved. 
Similarly, due to the fact that both $\mathcal{F}^\text{TD}$ and 
\begin{equation}
	{{P}_{{{{\mathbf{\tilde{Y}}}}_{i}}|{{\Lambda }_{i}}=\sqrt{{{\lambda }_{i}}}}}=\prod\limits_{j=1}^{n} {\mathcal{CN}\left(0,{1+\frac{\rho}{L}{{\lambda }_{i}}}\right)} \nonumber
\end{equation}
have spherical symmetry, $\beta_{1-\varepsilon,i}^n\left({{P}_{{{{\mathbf{\tilde{Y}}}}_{i}}|{{{\mathbf{\tilde{X}}}}_{i}}={{{\mathbf{\tilde{x}}}}_{i}},{{\Lambda }_{i}}=\sqrt{{{\lambda }_{i}}}}},{{P}_{{{{\mathbf{\tilde{Y}}}}_{i}}|{{\Lambda }_{i}}=\sqrt{{{\lambda }_{i}}}}}\right)$ does not depend on the codeword ${{{\mathbf{\tilde{x}}}}_{i}}\in \mathcal{F}^\text{TD}$. 
Therefore, we choose the codeword as ${{{\mathbf{\tilde{x}}}}_{i}}=\left(\sqrt{\rho/L},\dots,\sqrt{\rho/L}\right)^\text{T}$, at which point the information density can be obtained as
\begin{align}
	&{{\iota}_{i}}\left( {{{\mathbf{\tilde{x}}}}_{i}};{{{\mathbf{\tilde{Y}}}}_{i}}|{{\lambda }_{i}}\right)\nonumber \\
	&\quad=\sum\limits_{j=1}^{n}{\left[ \log \left( 1+\frac{\rho }{L}{{\lambda }_{i}} \right)+\left( 1-\frac{{{\left| \sqrt{\frac{\rho }{L}{{\lambda }_{i}}}{{W}_{i,j}}-1 \right|}^{2}}}{1+\frac{\rho }{L}{{\lambda }_{i}}} \right)\log e \right]}.
\end{align}

\begin{proposition}
	Using the Berry-Esseen CLT, the Neyman-Pearson function for link $i$ in the MIMO channel under time-domain coding can be expressed as
	\begin{align}
		&\beta_{1-\varepsilon,i}^n\left({{P}_{{{{\mathbf{\tilde{Y}}}}_{i}}|{{{\mathbf{\tilde{X}}}}_{i}}={{{\mathbf{\tilde{x}}}}_{i}},{{\Lambda }_{i}}=\sqrt{{{\lambda }_{i}}}}},{{P}_{{{{\mathbf{\tilde{Y}}}}_{i}}|{{\Lambda }_{i}}=\sqrt{{{\lambda }_{i}}}}}\right)\nonumber \\
		&\qquad=-nC_{i}^\text{TD}\left( \mathsf{H} \right) + \sqrt{nV_{i}^\text{TD}\left( \mathsf{H} \right)}\Phi^{-1}\left(\varepsilon\right)+\mathcal{O}\left(\log n\right),
	\end{align}
	where in a channel use, the channel capacity and channel dispersion of link $i$ are respectively
	\begin{align}
		C_{i}^\text{TD}\left( \mathsf{H} \right)&=\frac{1}{n}\mathbb{E}\left[ {{\iota}_{i}}\left( {{{\mathbf{\tilde{x}}}}_{i}};{{{\mathbf{\tilde{Y}}}}_{i}}|{{\lambda }_{i}} \right) \right]=\log \left( 1+\frac{\rho }{L}{{\lambda }_{i}} \right), \\
		V_{i}^\text{TD}\left( \mathsf{H} \right)&=\frac{1}{n}\text{Var}\left[ {{\iota}_{i}}\left( {{{\mathbf{\tilde{x}}}}_{i}};{{{\mathbf{\tilde{Y}}}}_{i}}|{{\lambda }_{i}} \right) \right]=\left[ 1-\frac{1}{{{\left( 1+\frac{\rho }{L}{{\lambda }_{i}} \right)}^{2}}} \right]{{\log }^{2}}e.
	\end{align}
	
	{\emph{Proof: }} The proof proceeds in the same way as Proposition 1, and can be considered as the case with one spatial DoF.
\end{proposition}

\begin{remark}
The SVD for the information density analysis of the equivalent channel is not in inconsistent with the assumption of unknown CSI at the transmitter.
We can consider that the matrices $\mathsf{U}$, $\mathsf{\Lambda}$, and $\mathsf{V}$ after SVD are all unknown at the transmitter, in which case the transmitter does not rely on CSI information for power allocation. 
Simultaneously, according to the spherical symmetry of the Neyman-Pearson function, the same result can be achieved by selecting any codeword that satisfies the codeword constraint.
This implies that it is unnecessary to know the matrix information of SVD at the transmitter, and only the codewords convenient for analysis need to be selected. 
\end{remark}

In the following, we analyze the different bounds and normal approximation for each independent link, respectively.

\subsubsection{Achievability}
We again assume $\tau=B/\sqrt{n}$ (here $B$ has the same definition as in spatiotemporal coding), in which case the achievability bound on the maximal number of codewords for link $i$ is
\begin{align} \label{ache2}
	&\log M_{i}^{*}\left( n,\varepsilon  \right) \nonumber \\
	&\ge \log {\kappa_\tau^n\left(\mathcal{F}^\text{TD},{{P}_{{{{\mathbf{\tilde{Y}}}}_{i}}|{{\Lambda }_{i}}=\sqrt{{{\lambda }_{i}}}}}\right)}\nonumber \\
	&\quad-\log{\beta_{1-\varepsilon+\tau,i}^n\left( {{P}_{{{{\mathbf{\tilde{Y}}}}_{i}}|{{{\mathbf{\tilde{X}}}}_{i}}={{{\mathbf{\tilde{x}}}}_{i}},{{\Lambda }_{i}}=\sqrt{{{\lambda }_{i}}}}},{{P}_{{{{\mathbf{\tilde{Y}}}}_{i}}|{{\Lambda }_{i}}=\sqrt{{{\lambda }_{i}}}}} \right)} \nonumber \\
	&=n{{C}_i^\text{TD}}\left( \mathsf{H} \right)-\sqrt{n{{V}_i^\text{TD}}\left( \mathsf{H} \right)}\Phi^{-1}\left(\varepsilon-\tau \right) +\log \kappa_\tau^n+ \mathcal{O}\left(\log n\right) \nonumber \\
	&\overset{\left( a \right)}{\mathop{\ge }} n{{C}_i^\text{TD}}\left( \mathsf{H} \right)-\sqrt{n{{V}_i^\text{TD}}\left( \mathsf{H} \right)}\Phi^{-1}\left(\varepsilon \right) + \mathcal{O}\left(\log n\right).
\end{align}
where $\left(a\right)$ is by Taylor's theorem and Lemma \ref{lemma6}.

\subsubsection{Converse}
For every $M_{i}^{*}\left( n,\varepsilon  \right)$-code belonging to $\mathcal{F}^\text{TD}$, the converse bound can be solved by Lemma \ref{lemma4} as
\begin{align}\label{con2}
	\log M_i^{*}\left( n,\varepsilon  \right) &\le -\log{\beta_{1-\varepsilon,i}^n\left( {{P}_{{{{\mathbf{\tilde{Y}}}}_{i}}|{{{\mathbf{\tilde{X}}}}_{i}}={{{\mathbf{\tilde{x}}}}_{i}},{{\Lambda }_{i}}=\sqrt{{{\lambda }_{i}}}}},{{P}_{{{{\mathbf{\tilde{Y}}}}_{i}}|{{\Lambda }_{i}}=\sqrt{{{\lambda }_{i}}}}} \right)} \nonumber \\
	&= n{{C}_i^\text{TD}}\left( \mathsf{H} \right)-\sqrt{n{{V}_i^\text{TD}}\left( \mathsf{H} \right)}\Phi^{-1}\left(\varepsilon \right) + \mathcal{O}\left(\log n\right).
\end{align}

\subsubsection{Normal Approximation}
By combining (\ref{ache2}) and (\ref{con2}), the maximal number of codewords $M_{i}^{*}\left( n,\varepsilon  \right)$ of link $i$ meets the requirement
\begin{equation}
\log M_{i}^{*}\left( n,\varepsilon  \right)\approx nC_{i}^\text{TD}\left( \mathsf{H} \right)-\sqrt{nV_{i}^\text{TD}\left( \mathsf{H} \right)}{{\Phi }^{-1}}\left( \varepsilon  \right).
\end{equation}

If $\varepsilon $ is small, there is $1-{{\left( 1-\varepsilon  \right)}^{m}}\approx m\varepsilon $, and if $m$ is also relatively small \footnote{According to numerical tests, under the typical parameter setting of $\varepsilon ={{10}^{-7}}$, the relative error is 5\% when $m=4$ and 19\% when $m=64$.}, there is ${{\Phi }^{-1}}\left( \varepsilon  \right)\approx {{\Phi }^{-1}}\left( m\varepsilon  \right)$. 
Therefore, for equivalent parallel time-domain coding of MIMO channel $\mathsf{H}$, the maximal number of codewords that can be achieved is
\begin{align}
	&\log M_\text{TD}^{*}\left( n,\varepsilon \right) \approx \log M_\text{TD}^{*}\left( n,1-{{\left( 1-\varepsilon  \right)}^{m}} \right) \nonumber \\
	& =\sum\limits_{i=1}^{m}{\log M_{i}^{*}\left( n,\varepsilon  \right)} \nonumber \\ 
	& \approx n\sum\limits_{i=1}^{m}{C_{i}^\text{TD}\left( \mathsf{H} \right)}-\sum\limits_{i=1}^{m}{\sqrt{nV_{i}^\text{TD}\left( \mathsf{H} \right)}}{{\Phi }^{-1}}\left( \varepsilon  \right) \nonumber \\ 
	& =n\sum\limits_{i=1}^{m}{\log \left( 1+\frac{\rho }{L}{{\lambda }_{i}} \right)}-\sum\limits_{i=1}^{m}{\sqrt{n\left[ 1-\frac{1}{{{\left( 1+\frac{\rho }{L}{{\lambda }_{i}} \right)}^{2}}} \right]}}\frac{{{\Phi }^{-1}}\left( \varepsilon  \right)}{\ln 2}.
\end{align}
Therefore, the channel capacity and channel dispersion of MIMO channel $\mathsf{H}$ using time-domain coding can be expressed as
\begin{align}
{{C}^\text{TD}}\left( \mathsf{H} \right)&=\sum\limits_{i=1}^{m}{\log \left( 1+\frac{\rho }{L}{{\lambda }_{i}} \right)}=\sum\limits_{i=1}^{m}{{{C}_{i}}}, \label{c3} \\
{{V}^\text{TD}}\left( \mathsf{H} \right)&={{\left[ \sum\limits_{i=1}^{m}{\sqrt{1-\frac{1}{{{\left( 1+\frac{\rho }{L}{{\lambda }_{i}} \right)}^{2}}}}} \right]}^{2}}{{\log }^{2}}e={{\left( \sum\limits_{i=1}^{m}{\sqrt{{{V}_{i}}}} \right)}^{2}}. \label{v3}
\end{align}

Then, for the channel matrix $\mathsf{H}$, the maximal coding rate should be
\begin{align}\label{r3}
	R_\text{TD}^{*}\left( n,\varepsilon  \right) &= \frac{\log M_\text{TD}^{*}\left( n,\varepsilon  \right)}{n} \nonumber \\
	& \approx {{C}^\text{TD}}\left( \mathsf{H} \right)-\sqrt{\frac{{{V}^\text{TD}\left( \mathsf{H} \right)}}{n}}{{\Phi }^{-1}}\left( \varepsilon  \right).
\end{align}

\begin{remark}
In contrast to spatiotemporal coding, the accuracy of the normal approximation in the case of time-domain coding depends on a single link. 
As a result, an increment in the spatial DoF fails to enhance the accuracy of the normal approximation for time-domain coding. 
However, it can still be regarded as accurate under circumstances such as an increase in blocklength, a requisite reduction in decoding error probability, or a high SNR.
Consequently, under our high SNR scenario assumption, it is reasonable to consider that the utilization of the normal approximation for analyzing the performance of time-domain coding is also accurate as described in Theorem \ref{the1}.
\end{remark}

\subsubsection{Maximal Achievable Rate}
In the case of the MIMO random channel matrix $\mathbb{H}$ and under the condition that $\rho/L\gg 1$, we also need to calculate the expectation of the channel capacity and channel dispersion in order to obtain the ergodic result of the maximal achievable rate in time-domain coding.
For channel capacity, the results of time-domain coding and spatiotemporal coding are precisely the same.
Thus, we can straightforwardly obtain that
\begin{equation}\label{c4}
\mathbb{E}\left[ {{C}^\text{TD}}\left( \mathbb{H} \right) \right]=\mathbb{E}\left[ \sum\limits_{i=1}^{m}{\log \left( 1+\frac{\rho }{L}{{\lambda }_{i}} \right)} \right]\approx m\log \left( 1+\rho  \right).
\end{equation}
The expectation of channel dispersion for time-domain coding can be obtained from Appendix \ref{ap3}:
\begin{equation}\label{v4}
\mathbb{E}\left[ \sqrt{{{V}^\text{TD}}\left( \mathbb{H} \right)} \right]=\mathbb{E}\left[ \sum\limits_{i=1}^{m}{\sqrt{1-\frac{1}{{{\left( 1+\frac{\rho }{L}{{\lambda }_{i}} \right)}^{2}}}}}\log e \right]\approx m\log e.
\end{equation}
Therefore, the upper bound of the maximal achievable rate of the MIMO channel in the case of time-domain coding can be obtained as
\begin{align}\label{r4}
{{{\bar{R}}}_\text{TD}}& =\mathbb{E}\left[ {{C}^\text{TD}}\left( \mathbb{H} \right) \right]-\frac{\mathbb{E}\left[ \sqrt{{{V}^\text{TD}}\left( \mathbb{H} \right)} \right]}{\sqrt{n}}{{\Phi }^{-1}}\left( \varepsilon  \right) \nonumber \\ 
& \xrightarrow{{\rho }/{L}\;\gg 1}m\log \left( 1+\rho  \right)-\frac{m}{\sqrt{n}}\frac{{{\Phi }^{-1}}\left( \varepsilon  \right)}{\ln 2}  .
\end{align}
\begin{remark}
Since the MIMO channel with $m$ spatial DoFs can be equivalent to $m$ parallel orthogonal links and the channel capacity and channel dispersion pairs of these $m$ independent links are $\left( {{C}_{i}},{{V}_{i}} \right),i=1,\ldots m$, it can be determined that the channel capacity of the MIMO system remains consistent in the cases of both spatiotemporal coding and time-domain coding, which is expressed as $C=\sum\nolimits_{i=1}^{m}{{{C}_{i}}}$. However, the channel dispersion
$$
V=
\begin{cases}
	\sum\limits_{i=1}^{m}{{{V}_{i}}},  & \text{Spatiotemporal coding} \\
	{{\left( \sum\limits_{i=1}^{m}{\sqrt{{{V}_{i}}}} \right)}^{2}}, & \text{Time-domin coding}
\end{cases}
$$
Since ${{\left( \sum\nolimits_{i=1}^{m}{\sqrt{{{V}_{i}}}} \right)}^{2}}>\sum\nolimits_{i=1}^{m}{{{V}_{i}}}$, independent coding leads to an increment in channel dispersion and, consequently, a reduction in the maximal achievable rate of the MIMO system.
\end{remark}

\section{Performance Comparison Between Time-Domain Coding and Spatiotemporal Coding}
In this section, we conduct an analysis of the advantages of spatiotemporal coding over time-domain coding from the perspectives of normalized maximal achievable rate (i.e., the average maximal achievable rate per link) and decoding error probability, respectively.

\subsection{Normalized Maximal Achievable Rate}
Since MIMO channels can be transformed into multiple parallel orthogonal links after SVD, the average maximal achievable rate of all links plays an important role in the performance analysis of finite blocklength coded MIMO. 
For this reason, we divide the maximal achievable rate of the MIMO system obtained via time-domain coding and spatiotemporal coding, respectively, by the spatial DoF $m$ to acquire the normalized maximal achievable rate.

In the high per-antenna SNR regime, for time-domain coding,
\begin{equation}
\frac{{{{\bar{R}}}_\text{TD}}}{m}\approx \log \left( 1+\rho  \right)-\frac{1}{\sqrt{n}}\frac{{{\Phi }^{-1}}\left( \varepsilon  \right)}{\ln 2}.
\end{equation}
It can be observed that, given the decoding error probability, the normalized maximal achievable rate of the MIMO system employing multi-antenna time-domain coding is the same as that of the single-antenna system under finite blocklength and will gradually decrease with the decrease of blocklength. 
This indicates that the utilization of time-domain coding in the MIMO system fails to exploit the spatial advantage of the MIMO system, thereby precluding the guarantee of the transmission performance of URLLC.

For spatiotemporal coding,
\begin{equation}\label{r_st}
\frac{{{{\bar{R}}}_\text{ST}}}{m}\approx \log \left( 1+\rho  \right)-\sqrt{\frac{1}{mn}}\frac{{{\Phi }^{-1}}\left( \varepsilon  \right)}{\ln 2}.
\end{equation}
It can be found that spatiotemporal coding exhibits a higher normalized maximal achievable rate under finite blocklength compared to time-domain coding, especially in the scenario of larger antenna arrays.
From the second term on the right-hand side of (\ref{r_st}), it is evident that spatiotemporal coding is capable of compensating for the performance degradation resulting from the progressively decreasing blocklength by increasing the spatial DoF, which is infeasible in time-domain coding.
This interesting phenomenon is called spatiotemporal exchangeability, that is, when the blocklength continuously diminishes, we can increase the spatial DoF and implement spatiotemporal coding to ensure the constancy of the coding rate and reliability.
Therefore, spatiotemporal coding can make full use of the spatial dimension advantage of MIMO to achieve extremely low latency communication.

\subsection{Decoding Error Probability}
For a given coding rate per link ${{\bar{R}}}/{m}$, the transmission reliability of the system can be reflected by decoding error probability $\varepsilon$.
For time-domain coding,
\begin{align}
	{{\varepsilon }^\text{TD}}& \approx \Phi \left[ \frac{\mathbb{E}\left[ {{C}^\text{TD}}\left( \mathbb{H} \right) \right]-\bar{R}}{{\mathbb{E}\left[ \sqrt{{{V}^\text{TD}}\left( \mathbb{H} \right)} \right]}/{\sqrt{n}}\;} \right] \nonumber \\ 
	& =\Phi \left[ \frac{m\log \left( 1+\rho  \right)-\bar{R}}{{m\log e}/{\sqrt{n}}\;} \right] \nonumber \\ 
	& =\Phi \left[ \left( \log \left( 1+\rho  \right)-\frac{{\bar{R}}}{m} \right)\sqrt{n}\ln 2 \right]  .
\end{align}
It can be observed that the decoding error probability exhibits a significant increase with the reduction in blocklength. 
Moreover, the spatial DoF does not contribute to the reduction of the decoding error probability in MIMO systems when using time-domain coding.

For spatiotemporal coding,
\begin{align}
	{{\varepsilon }^\text{ST}}& \approx \Phi \left[ \frac{\mathbb{E}\left[ {{C}^\text{ST}}\left( \mathbb{H} \right) \right]-\bar{R}}{{\sqrt{\mathbb{E}\left[ {{V}^\text{ST}}\left( \mathbb{H} \right) \right]}}/{\sqrt{n}}\;} \right] \nonumber \\ 
	& =\Phi \left[ \frac{m\log \left( 1+\rho  \right)-\bar{R}}{{\sqrt{m}\log e}/{\sqrt{n}}\;} \right]  \nonumber \\ 
	& =\Phi \left[ \left( \log \left( 1+\rho  \right)-\frac{{\bar{R}}}{m} \right)\sqrt{mn}\ln 2 \right]  .
\end{align}
It can be found that in the case of spatiotemporal coding, as the spatial DoF $m$ increase, the growth in decoding error probability induced by the continuous reduction of blocklength can be mitigated.
We define $\Delta =\log \left( 1+\rho  \right)-{{\bar{R}}}/{m}$, and when $mn\gg 1/{{\Delta }^{2}}$, it follows that ${{\varepsilon }^\text{ST}}\to 0$. 
This implies that in the case of spatiotemporal coding, even with a finite blocklength, an arbitrarily low decoding error probability can be attained by increasing the spatial DoF.
Therefore, from an alternative perspective, spatiotemporal coding can make full use of the spatial dimension advantages of MIMO to achieve high reliability in low-latency communication.

\begin{remark}
Spatiotemporal coding exhibits no additional coding delay in comparison with time-domain coding. 
Since spatiotemporal codewords are generated in a parallel manner, the output delay of the encoder is identical to that of the time-domain coding. 
Simultaneously, spatiotemporal codewords are also decoded in parallel during the decoding process.
Consequently, under the same code delay constraint, spatiotemporal coding is capable of enhancing the system performance compared with time-domain coding owing to the spatial DoF.
\end{remark}

\section{Simulation Results}
\begin{figure*}[ht]
	\centering
	\subfigure[$L=N=4$, $\rho = 0 \text{ dB}$]
	{
		\label{figure21}
		\begin{minipage}[b]{.31\linewidth}
			\centering
			\includegraphics[scale=0.42]{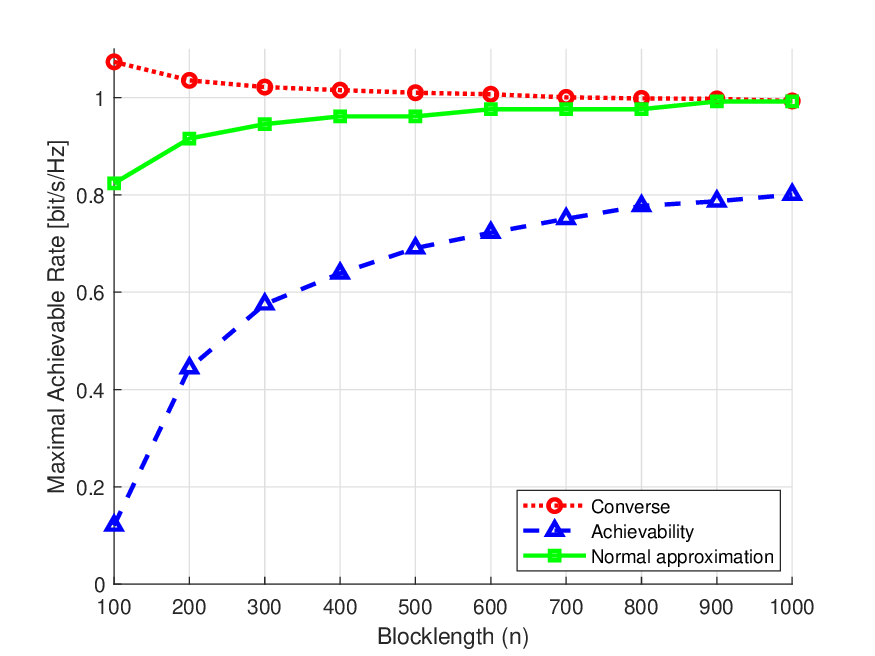}
		\end{minipage}
	}
	\subfigure[$L=N=4$, $\rho = 10 \text{ dB}$]
	{
		\label{figure22}
		\begin{minipage}[b]{.31\linewidth}
			\centering
			\includegraphics[scale=0.42]{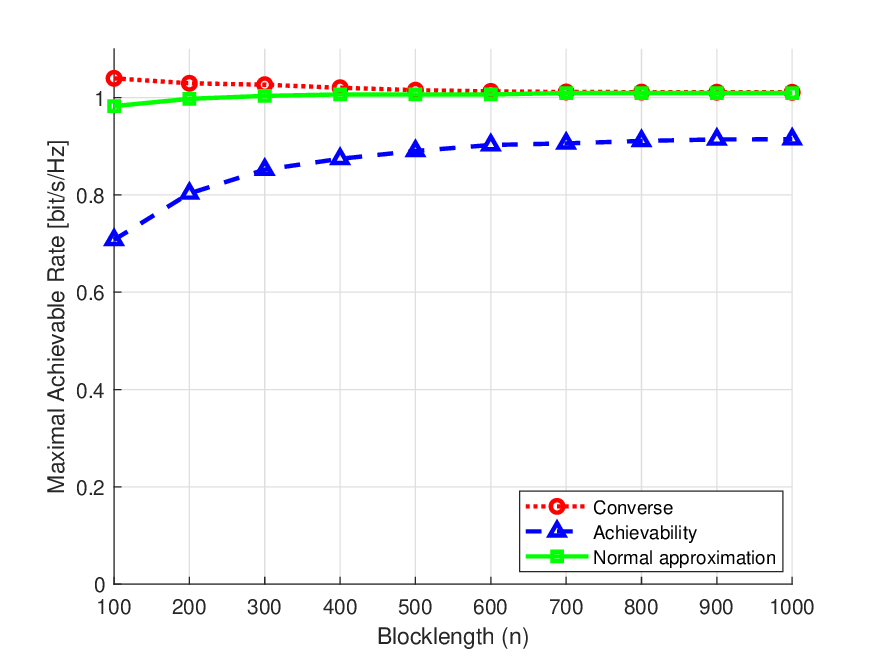}
		\end{minipage}
	}
	\subfigure[$L=N=16$, $\rho = 10 \text{ dB}$]
	{
		\label{figure23}
		\begin{minipage}[b]{.31\linewidth}
			\centering
			\includegraphics[scale=0.42]{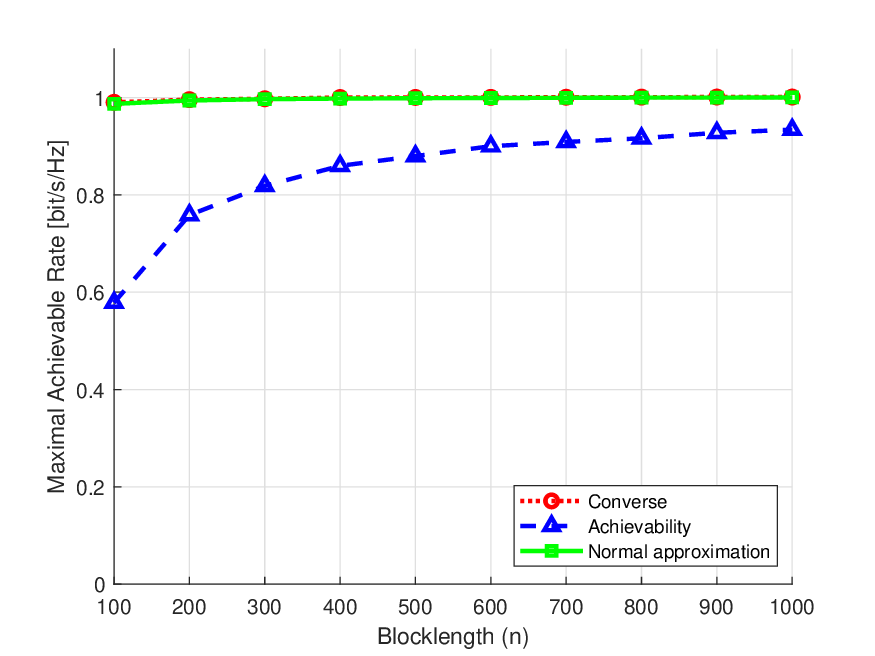}
		\end{minipage}
	}
	\caption{Comparison of the achievability and converse bounds with the normal approximation in spatiotemporal coding with $\varepsilon=10^{-7}$.}
	\label{figure2}
\end{figure*}

In this section, we initially verify the different bounds and normal approximation. 
Subsequently, we conduct a fitting of the simulation and approximation results of the expectation of channel dispersion and the normalized maximal achievable rates under diverse coding modes. 
Finally, the normalized maximal achievable rate and decoding error probability of spatiotemporal coding and time-domain coding are comparatively analyzed respectively.

We present Fig. \ref{figure2} in spatiotemporal coding as representatives for comparing the results of achievability (\ref{ache1}) and converse bounds (\ref{con1}) with the normal approximation (\ref{normal1}) in the case of the capacity is $1 \text{ bit/s/Hz}$. 
The only difference between the parameter settings in Fig. \ref{figure21} and Fig. \ref{figure22} lies in the SNR, which is higher in Fig. \ref{figure22}. 
Moreover, the only difference in the parameter settings between Fig. \ref{figure22} and Fig. \ref{figure23} is in the spatial DoF, which is larger in Fig. \ref{figure23}. 
We employ the distance between the normal approximation and the converse bound to characterize the accuracy of the normal approximation.
It is evidently discernible from the figures that an increment in SNR substantially improves the accuracy of the normal approximation, and the enhancement in accuracy is highly significant. 
The same holds true for the fact that the utilization of larger spatial DoF leads to a more accurate normal approximation.
Consequently, it is rational to characterize the maximal achievable system performance via the normal approximation in our high SNR massive MIMO scenario.
There is a distance between the performance of the $\kappa \beta$ bound in Lemma 3 and the normal approximation and the converse bound, indicating the requirement for better achievability bounds to characterize the performance of quasi-static MIMO fading channels.

\begin{figure}
	\centering
	\includegraphics[width=9cm]{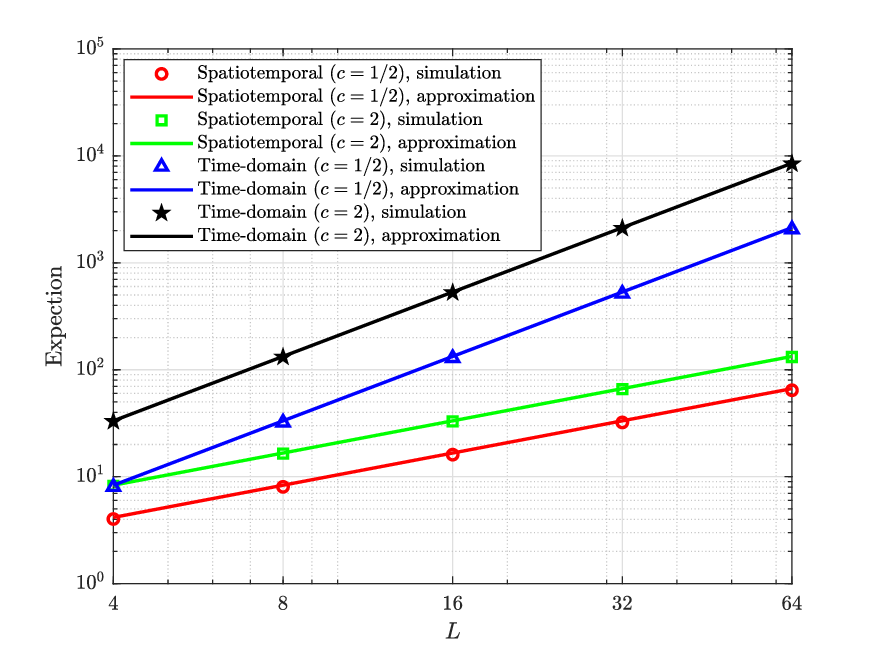}\\
	\caption{The fitting results of the expectation of channel dispersion under different coding modes with $\rho = 10 \text{ dB}$.}\label{figure5}
\end{figure}

Fig. \ref{figure5} illustrates the comparison of simulation and approximation values for the expectation of channel dispersion under different coding modes and different proportions of transceiver antennas, denoted as $c=L/N$.
The simulation results are derived from (\ref{v1}) and (\ref{v3}), while the approximation values are obtained from (\ref{v2}) and (\ref{v4}) (after deformation).
It is observable from the figure that the expectation of the channel dispersion is perfectly fitted under different coding methods and different transmit and receive antenna relationships.
In addition, the channel dispersion is determined by the minimum value in the number of transmit and receive antennas (i.e., the DoF). 
The channel dispersion is remarkably increased when time-domain coding is employed in contrast to spatiotemporal coding, especially in the case of a larger DoF.

\begin{figure}
	\centering
	\includegraphics[width=9cm]{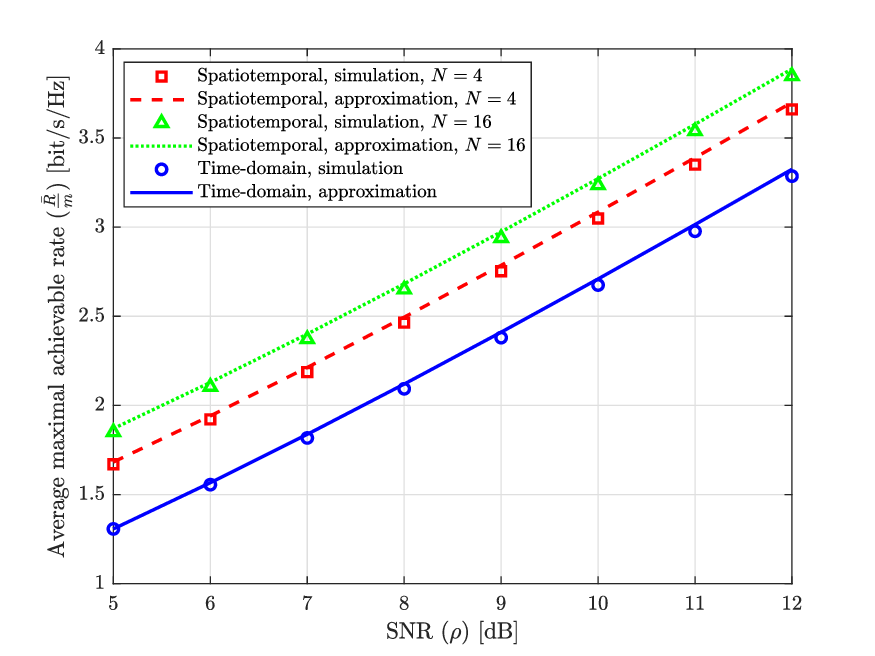}\\
	\caption{The fitting results of the average maximal achievable rate per link under different coding modes with fixed proportion of transceiver antennas $c=L/N=16$, $n=100$ and $\varepsilon={10}^{-7}$.}\label{figure6}
\end{figure}

Fig. \ref{figure6} presents the fitting results and comparison of the average maximal achievable rate per link under different coding modes and different numbers of receive antennas $N$ with the change of SNR.
The simulation results are obtained from (\ref{r1}) and (\ref{r3}), while the approximation values are derived from (\ref{r2}) and (\ref{r4}).
It can be observed from the figure that the average maximal achievable rate per link is satisfactorily fitted under different coding modes.
When the spatial DoF is greater than 1, the average maximal achievable rate per link of spatiotemporal coding is greater than that of time-domain coding, and the performance of spatiotemporal coding becomes better with the increase of spatial DoF. 
In addition, the average maximal achievable rate per link exhibits an approximately linear increase with the increase in SNR under different coding modes.

\begin{figure}
	\centering
	\includegraphics[width=9cm]{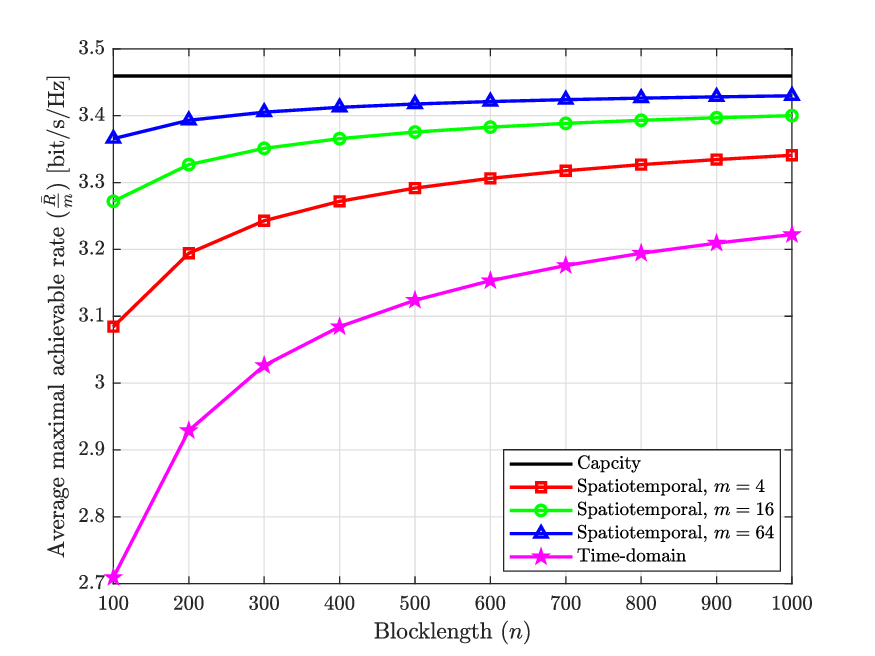}\\
	\caption{The relationship between the average maximal achievable rate per link and Shannon capacity under different coding modes with $\rho = 10 \text{ dB}$ and $\varepsilon={10}^{-7}$.}\label{figure7}
\end{figure}

Fig. \ref{figure7} depicts the variation of the average maximal achievable rate per link of spatiotemporal coding and time-domain coding with respect to blocklength under different spatial DoFs. 
As can be discerned from the figure, for a given blocklength, the average maximal achievable rate per link in spatiotemporal coding is more proximate to the Shannon capacity compared to that in time-domain coding.
Furthermore, with the continuous augmentation of spatial DoF, spatiotemporal coding can asymptotically approach the Shannon capacity.
In addition, under the given decoding error probability, the average maximal achievable rate per link undergoes a significant decrease as the blocklength diminishes.
This rate degradation can be substantially mitigated by increasing the spatial DoF and implementing spatiotemporal coding, whereas it cannot be alleviated by utilizing time-domain coding. 
This thereby demonstrates the indispensability of spatiotemporal coding in MIMO systems in the finite blocklength regime.

\begin{figure}
	\centering
	\includegraphics[width=9cm]{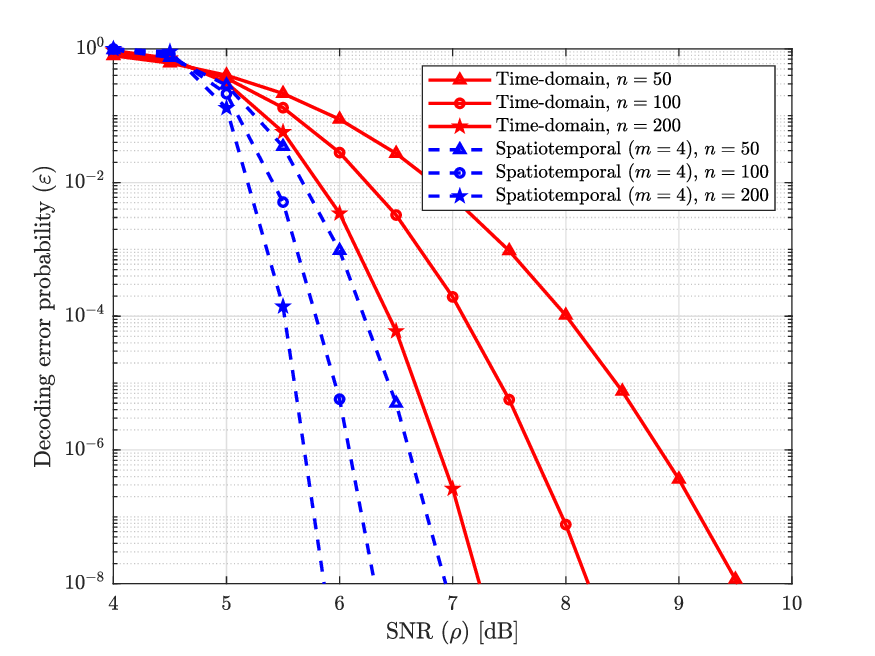}\\
	\caption{Comparison of decoding error probability under different coding modes and different blocklengths.}\label{figure8}
\end{figure}

Fig. \ref{figure8} presents a comparison of the decoding error probability of different coding methods and different blocklengths under the coding rate setting of $\bar{R}=2m $ bit/s/Hz, with the spatial DoF set as $m=4$. 
As can be observed from the figure, for a specific coding rate, when the spatial DoF exceeds 1, spatiotemporal coding can attain a higher level of reliability compared to time-domain coding. 
This verifies that spatiotemporal coding in MIMO systems is capable of achieving highly reliable communication in low-latency systems.
At the same time, to fulfill the same decoding error probability requirement (as illustrated by the time-domain coding with $n=200$ and the spatiotemporal coding with $n=50$ in the figure), the latency can be reduced by 4 times in spatiotemporal coding.
It is revealed that the latency can be further diminished and low-latency communication can be realized through spatiotemporal coding in MIMO systems.

\section{Conclusion}
In this paper, through an analysis of the normal approximation forms and statistical characteristics of information density, compact and explicit performance bounds of finite blocklength coded MIMO are established to investigate the relationship among blocklength, decoding error probability, rate, and spatial DoF in both time-domain coding and spatiotemporal coding.
It is demonstrated that spatiotemporal coding exhibits greater advantages over time-domain coding in finite blocklength coded MIMO with respect to reliability and latency, owing to the spatial DoF.
Furthermore, the results suggest that spatiotemporal coding holds the potential to achieve extreme connectivity, which demands both ultra-low latency and ultra-high reliability simultaneously.
We anticipate that these finite blocklength coding studies will offer novel tools for practical spatiotemporal code design in 6G MIMO URLLC.

\appendices
\section{Proof of Proposition 1}\label{ap1}
According to (\ref{ht1}), we have
\begin{align}
&\log {{\beta }^n_{\alpha }}\left( {{P}_{{\mathbb{Y}}|{\mathbb{X}=\mathsf{X},\mathbb{H}=\mathsf{H}}}},{{P}_{{\mathbb{Y}}|\mathbb{H}=\mathsf{H}}}\right)\nonumber \\
&\qquad\ge \frac{1}{\gamma }\left( \alpha -P\left[ \frac{d{{P}_{{\mathbb{Y}}|{\mathbb{X}=\mathsf{X},\mathbb{H}=\mathsf{H}}}}}{d{{P}_{{\mathbb{Y}}|\mathbb{H}=\mathsf{H}}}}\ge \gamma  \right] \right) \nonumber \\
&\qquad=\frac{1}{\gamma }\left( \alpha -P\left[\log \frac{d{{P}_{{\mathbb{Y}}|{\mathbb{X}=\mathsf{X},\mathbb{H}=\mathsf{H}}}}}{d{{P}_{{\mathbb{Y}}|\mathbb{H}=\mathsf{H}}}}\ge \log \gamma  \right] \right) \nonumber \\
&\qquad=\frac{1}{\gamma }\left( \alpha -P\left[\iota\left( \mathsf{X};\mathbb{Y}|\mathsf{H} \right)\ge \log \gamma  \right] \right).
\end{align}
Since the  capacity achieving input codewords are i.i.d, the input symbols at each channel use are independent of each other. Also, considering the assumption of a memoryless channel, we define $T=\mathbb{E}\left[ {{\left|\frac{1}{n} \iota\left( \mathsf{X};\mathbb{Y}|\mathsf{H} \right)-{{C}^\text{ST}}\left(\mathsf{H} \right) \right|}^{3}} \right]$, $B=6T/\left({{{V}^\text{ST}}\left( \mathsf{H} \right)}\right)^{\frac{3}{2}}$, $\Delta>0$ is a constant and $\log \gamma=n{{C}^\text{ST}}\left(\mathsf{H} \right)+\sqrt{n{{V}^\text{ST}}\left(\mathsf{H} \right)}\Phi^{-1}\left(\alpha-\frac{B+\Delta}{\sqrt{n}}\right)$, then we use the Berry-Esseen CLT and have
\begin{align}
&P\left[\iota\left( \mathsf{X};\mathbb{Y}|\mathsf{H} \right)\ge \log \gamma  \right]\nonumber \\
&\qquad=P\left[\frac{\iota\left( \mathsf{X};\mathbb{Y}|\mathsf{H} \right)-n{{C}^\text{ST}}\left(\mathsf{H} \right)}{\sqrt{n{{V}^\text{ST}}\left(\mathsf{H} \right)}}\ge \Phi^{-1}\left(\alpha-\frac{B+\Delta}{\sqrt{n}}\right) \right] \nonumber \\
&\qquad \le \Phi\left(\Phi^{-1}\left(\alpha-\frac{B+\Delta}{\sqrt{n}}\right)\right)+\frac{B}{\sqrt{n}} \nonumber \\
&\qquad = \alpha - \frac{\Delta}{\sqrt{n}} .
\end{align}
Therefore,
\begin{align}\label{bf1}
& \log {{\beta }^n_{\alpha }}\left( {{P}_{{\mathbb{Y}}|{\mathbb{X}=\mathsf{X},\mathbb{H}=\mathsf{H}}}},{{P}_{{\mathbb{Y}}|\mathbb{H}=\mathsf{H}}}\right)\nonumber \\
&\qquad\ge -\log \gamma + \log \Delta - \frac{1}{2} \log n \nonumber \\
&\qquad= -n{{C}^\text{ST}}\left(\mathsf{H} \right)- \sqrt{n{{V}^\text{ST}}\left(\mathsf{H} \right)}\Phi^{-1}\left(\alpha-\frac{B+\Delta}{\sqrt{n}}\right)\nonumber \\
&\qquad \quad + \log \Delta - \frac{1}{2} \log n \nonumber \\
&\qquad \overset{\left( a \right)}{\mathop{=}} -n{{C}^\text{ST}}\left(\mathsf{H} \right)- \sqrt{n{{V}^\text{ST}}\left(\mathsf{H} \right)}\Phi^{-1}\left(\alpha\right)+ \mathcal{O}\left(\log n\right) \nonumber \\
&\qquad \overset{\left( b \right)}{\mathop{=}} -n{{C}^\text{ST}}\left(\mathsf{H} \right)+ \sqrt{n{{V}^\text{ST}}\left(\mathsf{H} \right)}\Phi^{-1}\left(1-\alpha\right)+ \mathcal{O}\left(\log n\right),
\end{align}
where $\left(a\right)$ is by Taylor's theorem and $\left(b\right)$ is established because $\Phi^{-1}\left(1-\alpha\right)=-\Phi^{-1}\left(\alpha\right)$.

Next, we let $\gamma_0=n{{C}^\text{ST}}\left(\mathsf{H} \right)+\sqrt{n{{V}^\text{ST}}\left(\mathsf{H} \right)}\Phi^{-1}\left(\alpha+\frac{B}{\sqrt{n}}\right)$. Then, using the Berry-Esseen CLT in (\ref{ht3}), we can have
\begin{align}
&P\left[\iota\left( \mathsf{X};\mathbb{Y}|\mathsf{H} \right)\ge \log \gamma_0  \right]\nonumber \\
&\qquad=P\left[\frac{\iota\left( \mathsf{X};\mathbb{Y}|\mathsf{H} \right)-n{{C}^\text{ST}}\left(\mathsf{H} \right)}{\sqrt{n{{V}^\text{ST}}\left(\mathsf{H} \right)}}\ge \Phi^{-1}\left(\alpha+\frac{B}{\sqrt{n}}\right) \right] \nonumber \\
&\qquad \ge \Phi\left(\Phi^{-1}\left(\alpha+\frac{B}{\sqrt{n}}\right)\right)-\frac{B}{\sqrt{n}} \nonumber \\
&\qquad = \alpha.
\end{align}
This proves that the choice of $\gamma_0$ is correct.
Then, bring the result into (\ref{ht2}) to get
\begin{align}\label{bf2}
&\log {{\beta }_{\alpha }}\left( {{P}_{{\mathbb{Y}}|{\mathbb{X}=\mathsf{X},\mathbb{H}=\mathsf{H}}}},{{P}_{{\mathbb{Y}}|\mathbb{H}=\mathsf{H}}} \right)\nonumber \\
&\qquad\le -\log \gamma_0 \nonumber \\
&\qquad= -n{{C}^\text{ST}}\left(\mathsf{H} \right)- \sqrt{n{{V}^\text{ST}}\left(\mathsf{H} \right)}\Phi^{-1}\left(\alpha+\frac{B}{\sqrt{n}}\right) \nonumber \\
&\qquad \overset{\left( a \right)}{\mathop{=}} -n{{C}^\text{ST}}\left(\mathsf{H} \right)- \sqrt{n{{V}^\text{ST}}\left(\mathsf{H} \right)}\Phi^{-1}\left(\alpha\right)+ \mathcal{O}\left(\log n\right) \nonumber \\
&\qquad = -n{{C}^\text{ST}}\left(\mathsf{H} \right)+ \sqrt{n{{V}^\text{ST}}\left(\mathsf{H} \right)}\Phi^{-1}\left(1-\alpha\right)+ \mathcal{O}\left(\log n\right),
\end{align}
where $\left(a\right)$ also uses Taylor's theorem.

Finally, combining (\ref{bf1}) and (\ref{bf2}), we can finally get (\ref{ht4}).

\section{Proof of Theorem 1}\label{ap2}
According to the derivation procedure of the normal approximation for a quasi-static flat Rayleigh fading MIMO channel in (\ref{normal1}), when the higher order terms are small enough to be neglected, the normal approximation that preserves the first two order terms is accurate. Since the decay of the higher order terms is going to be less than $\log n$, we just need to figure out when
\begin{equation}\label{acc1}
n{{C}^\text{ST}}\left( \mathsf{H} \right)-\sqrt{n{{V}^\text{ST}}\left( \mathsf{H} \right)}\Phi^{-1}\left(\varepsilon \right) \gg \log n.
\end{equation}
In the subsequent part, we demonstrate the conditions under which (\ref{acc1}) holds by considering the following four aspects, respectively.

\subsection{Large Blocklength}
Since ${{C}^\text{ST}}\left( \mathsf{H} \right)$ and ${{V}^\text{ST}}\left( \mathsf{H} \right)$ are independent of $n$, the left half of (\ref{acc1}) can be transformed into
\begin{align}
&n{{C}^\text{ST}}\left( \mathsf{H} \right)-\sqrt{n{{V}^\text{ST}}\left( \mathsf{H} \right)}\Phi^{-1}\left(\varepsilon \right)\nonumber \\
&\qquad=n\left[{{C}^\text{ST}}\left( \mathsf{H} \right)-\sqrt{\frac{{{V}^\text{ST}}\left( \mathsf{H} \right)}{n}}\Phi^{-1}\left(\varepsilon \right)\right].
\end{align}
Therefore, to prove (\ref{acc1}) is to prove 
\begin{equation}\label{acc2}
{{C}^\text{ST}}\left( \mathsf{H} \right)-\sqrt{\frac{{{V}^\text{ST}}\left( \mathsf{H} \right)}{n}}\Phi^{-1}\left(\varepsilon \right) \gg \frac{\log n}{n}.
\end{equation}
It can be readily observed that the left-hand side of (\ref{acc2}) exhibits an increasing trend as $n$ increases.
Regarding the right-hand side of (\ref{acc2}), if we define $f\left(n\right)=\frac{\log n}{n}$, then we have
\begin{equation}
{f}'\left( n \right)=\frac{\frac{1}{\ln 2}-\log x}{{{x}^{2}}}.
\end{equation}
When the derivative ${f}'\left( n \right)>0$, i.e., $n>e$, the function $f\left(n\right)$ is a monotonically decreasing function.
In practical scenarios, the condition $n>e$ is invariably satisfied. 
This implies that the right-hand side of (\ref{acc2}) decreases as $n$ increases. 
In summary, as $n$ increases, the inequality (\ref{acc1}) can hold, which indicates that a larger blocklength has the potential to enhance the accuracy of the normal approximation.

\subsection{High Decoding Error Probability}
Since $\Phi^{-1}\left(\varepsilon \right)$ is a monotonically decreasing function, it follows that the left-hand side of (\ref{acc1}) will attain larger values for greater $\varepsilon$. 
Consequently, the formulation of the normal approximation will exhibit enhanced accuracy when the decoding error probability is higher.

\subsection{High SNR}
We take the derivative of ${{C}^\text{ST}}\left( \mathsf{H} \right)$ and ${{V}^\text{ST}}\left( \mathsf{H} \right)$ with respect to $\rho$ and get that
\begin{align}
\frac{\partial {{C}^{\text{ST}}}\left( \mathsf{H} \right)}{\partial \rho }&=\sum\limits_{i=1}^{m}{\frac{{{\lambda }_{i}}}{L\left( 1+\frac{\rho {{\lambda }_{i}}}{L} \right)}}>0, \\
\frac{\partial {{V}^{\text{ST}}}\left(\mathsf{H} \right)}{\partial \rho }&=\sum\limits_{i=1}^{m}{\frac{2\lambda _{i}^{2}}{{{L}^{2}}{{\left( 1+\frac{\rho {{\lambda }_{i}}}{L} \right)}^{3}}}}{{\log }^{2}}e>0.
\end{align}
It can be observed that both ${{C}^\text{ST}}\left( \mathsf{H} \right)$ and ${{V}^\text{ST}}\left( \mathsf{H} \right)$ exhibit an increasing behavior as $\rho$ increases. 
Simultaneously, considering that $\frac{\partial {{C}^{\text{ST}}}\left( \mathsf{H} \right)}{\partial \rho }$ is inversely proportional to $\rho$, and $\frac{\partial {{V}^{\text{ST}}}\left(\mathsf{H} \right)}{\partial \rho }$ is inversely proportional to the cube of $\rho$, it can be inferred that as $\rho$ rises, the rate of increase of ${{C}^{\text{ST}}}\left( \mathsf{H} \right)$ is substantially faster than that of ${{V}^{\text{ST}}}\left( \mathsf{H} \right)$. 
Consequently, when $\rho$ increases, the rate of increase of ${{C}^{\text{ST}}}\left( \mathsf{H} \right)$ is even more evidently faster than that of $\sqrt{{{V}^{\text{ST}}}\left( \mathsf{H} \right)}$.
Meanwhile, given that the inequality $n{{C}^{\text{ST}}}\left( \mathsf{H} \right)>\sqrt{n{{V}^{\text{ST}}}\left( \mathsf{H} \right)}\Phi^{-1}\left(\varepsilon \right)$ holds constantly, it can be deduced that $n{{C}^{\text{ST}}}\left( \mathsf{H} \right)-\sqrt{n{{V}^{\text{ST}}}\left( \mathsf{H} \right)}\Phi^{-1}\left(\varepsilon \right)$ will increase with the augmentation of $\rho$. 
This indicates that when the SNR is considerably high, $n{{C}^\text{ST}}\left( \mathsf{H} \right)-\sqrt{n{{V}^\text{ST}}\left( \mathsf{H} \right)}\Phi^{-1}\left(\varepsilon \right) \gg \log n$.
Therefore, when a high SNR is utilized, the normal approximation can precisely characterize the maximal achievable performance of the system.

\subsection{Large Spatial DoF}
Let ${{C}_m^{\text{ST}}}\left( \mathsf{H} \right)={{C}^{\text{ST}}}\left( \mathsf{H} \right)$ and ${{V}_m^{\text{ST}}}\left( \mathsf{H} \right)={{V}^{\text{ST}}}\left( \mathsf{H} \right)$.
When the spatial DoF is incremented by one, we have
\begin{align}
{{C}_{m+1}^{\text{ST}}}\left( \mathsf{H} \right)&=\sum\limits_{i=1}^{m+1}{\log \left( 1+\frac{\rho }{L}{{\lambda }_{i}} \right)}, \\
{{V}_{m+1}^{\text{ST}}}\left( \mathsf{H} \right)&=\sum\limits_{i=1}^{m+1}{\left[ 1-\frac{1}{{{\left( 1+\frac{\rho }{L}{{\lambda }_{i}} \right)}^{2}}} \right]}{{\log }^{2}}e.
\end{align}
In order to prove (\ref{acc1}), specifically, to demonstrate that $n{{C}^{\text{ST}}}\left( \mathsf{H} \right)-\sqrt{n{{V}^{\text{ST}}}\left( \mathsf{H} \right)}$ increases as $m$ increases, it is sufficient to prove that
\begin{align}\label{dofm1}
&n{{C}_{m+1}^{\text{ST}}}\left( \mathsf{H} \right)-\sqrt{n{{V}_{m+1}^{\text{ST}}}\left( \mathsf{H} \right)}\Phi^{-1}\left(\varepsilon \right)\nonumber \\
&\qquad-\left[n{{C}_{m}^{\text{ST}}}\left( \mathsf{H} \right)-\sqrt{n{{V}_{m}^{\text{ST}}}\left( \mathsf{H} \right)}\Phi^{-1}\left(\varepsilon \right)\right] \nonumber \\
&= n{\log \left( 1+\frac{\rho }{L}{{\lambda }_{m+1}} \right)}+\sqrt{n{{V}_{m}^{\text{ST}}}\left( \mathsf{H} \right)}\Phi^{-1}\left(\varepsilon \right)\nonumber \\
&\quad-\sqrt{n\left[{\left( 1-\frac{1}{{{\left( 1+\frac{\rho }{L}{{\lambda }_{m+1}} \right)}^{2}}} \right)\log^2e+{{V}_{m}^{\text{ST}}}\left( \mathsf{H} \right)}\right]}\Phi^{-1}\left(\varepsilon \right)\nonumber \\
&>0.
\end{align} 
Since the coding rate must be positive, i.e., 
\begin{equation}
n{\log \left( 1+\frac{\rho }{L}{{\lambda }_{m+1}} \right)} > \sqrt{n{\left( 1-\frac{1}{{{\left( 1+\frac{\rho }{L}{{\lambda }_{m+1}} \right)}^{2}}} \right)\log^2e}}\Phi^{-1}\left(\varepsilon \right)
\end{equation}
is established, we only need to show that
\begin{align}\label{ineq1}
	&\sqrt{n{\left( 1-\frac{1}{{{\left( 1+\frac{\rho }{L}{{\lambda }_{m+1}} \right)}^{2}}} \right)\log^2e}}+\sqrt{n{{V}_{m}^{\text{ST}}}\left( \mathsf{H} \right)} \nonumber \\
	&\quad > \sqrt{n\left[{\left( 1-\frac{1}{{{\left( 1+\frac{\rho }{L}{{\lambda }_{m+1}} \right)}^{2}}} \right)\log^2e+{{V}_{m}^{\text{ST}}}\left( \mathsf{H} \right)}\right]}.
\end{align}
Taking the square of both sides of (\ref{ineq1}) simultaneously, we found it holds true constantly.
This indicates that as the spatial DoF increases, $n{{C}^\text{ST}}\left( \mathsf{H} \right)-\sqrt{n{{V}^\text{ST}}\left( \mathsf{H} \right)}\Phi^{-1}\left(\varepsilon \right) \gg \log n$, and the normal approximation will become more accurate.

\section{Derivation of Channel Dispersion in Time-Domain Coding}\label{ap3}
For the expectation of channel dispersion in time-domain coding, we have
\begin{equation}
\mathbb{E}\left[ \sqrt{{{V}^\text{TD}}\left( \mathbb{H} \right)} \right]=\mathbb{E}\left[ \sum\limits_{i=1}^{m}{\sqrt{1-\frac{1}{{{\left( 1+\frac{\rho }{L}{{\lambda }_{i}} \right)}^{2}}}}} \right]\log e.
\end{equation}
Let ${{g}_{i}}=-\frac{1}{{{\left( 1+\frac{\rho }{L}{{\lambda }_{i}} \right)}^{2}}}$, then $\mathbb{E}\left[ \sqrt{{{V}^\text{TD}}\left( \mathbb{H} \right)} \right]=\mathbb{E}\left[ \sum\limits_{i=1}^{m}{{{\left( 1+{{g}_{i}} \right)}^{\frac{1}{2}}}} \right]\log e$.
Since ${{\lambda }_{i}}>0$ and ${\rho }/{L}\;\gg 1$, then $1+\frac{\rho }{L}{{\lambda }_{i}}>1$ and $-1<{{g}_{i}}<0$.
Taking a two-order Taylor expansion of ${{\left( 1+{{g}_{i}} \right)}^{\frac{1}{2}}}$, we can get ${{\left( 1+{{g}_{i}} \right)}^{\frac{1}{2}}}\approx 1+\frac{1}{2}{{g}_{i}}$.
Thus we have
\begin{align}
	\mathbb{E}\left[ \sum\limits_{i=1}^{m}{{{\left( 1+{{g}_{i}} \right)}^{\frac{1}{2}}}} \right]& =\mathbb{E}\left[ \sum\limits_{i=1}^{m}{\left( 1+\frac{1}{2}{{g}_{i}} \right)} \right] \nonumber \\ 
	& =\mathbb{E}\left[ \sum\limits_{i=1}^{m}{\left( 1-\frac{1}{2}\frac{1}{{{\left( 1+\frac{\rho }{L}{{\lambda }_{i}} \right)}^{2}}} \right)} \right] \nonumber \\ 
	& =m-\frac{1}{2}\mathbb{E}\left[ \sum\limits_{i=1}^{m}{\frac{1}{{{\left( 1+\frac{\rho }{L}{{\lambda }_{i}} \right)}^{2}}}} \right] \nonumber \\
	& \overset{(a)}{\mathop{\approx }}\,m-\frac{{{L}^{2}}}{2\rho }\mathbb{E}\left[ \sum\limits_{i=1}^{m}{\frac{1}{\lambda _{i}^{2}}} \right] \nonumber \\ 
	& \overset{(b)}{\mathop{=}}\,m-\frac{{{L}^{2}}}{2\rho }\frac{LN}{{{\left| L-N \right|}^{3}}-\left| L-N \right|} \nonumber \\ 
	& \overset{(c)}{\mathop{\approx }}\,m  ,
\end{align}
where the conditions for the establishment of (a) and (c) are per-antenna high SNR ${\rho }/{L}\;\gg 1$. 
The reason for the establishment of (b) is that we let
$$\mathbb{U}=\left\{ \begin{matrix}
	{{\mathbb{H}}^{\text{H}}}\mathbb{H},L\ge N  \\
	\mathbb{H}{{\mathbb{H}}^{\text{H}}},L<N  \\
\end{matrix} \right.$$
be a Wishart matrix, then according to \cite{tulino2004random} we have
\begin{equation}
\mathbb{E}\left[ \text{tr}\left( {{\mathbb{U}}^{-2}} \right) \right]=\frac{LN}{{{\left| L-N \right|}^{3}}-\left| L-N \right|}
\end{equation}

In summary, we can finally get (\ref{v4}).

\balance
\bibliographystyle{ieeetr}
\bibliography{ps}

\end{document}